\documentclass[a4paper,11pt,headings=big]{scrartcl}
\pdfoutput=1
\usepackage[utf8x]{inputenc}
\usepackage[colorlinks,linkcolor=darkblue,citecolor=darkblue,urlcolor=darkblue]{
hyperref}
\usepackage{graphicx}
\usepackage{tabularx}
\usepackage{dsfont, amsmath,amssymb, amsfonts,slashed}
\usepackage{array}
\usepackage{geometry}
\usepackage{xcolor}
\usepackage{longtable}
\usepackage{verbatim}
\usepackage{bbm}
\usepackage{multirow}
\usepackage{colortbl}
\usepackage{slashed}
\usepackage{float}
\usepackage[normalem]{ulem}
\definecolor{darkblue}{rgb}{0,0,0.5}
\allowdisplaybreaks

\newcommand{\T}{\mathsf{T}}
\newcommand{\im}{\mathrm{i}}
\newcommand{\U}{\mathcal{U}}
\newcommand{\sh}{s_h}
\newcommand{\ch}{c_h}
\newcommand{\seta}{s_\eta}
\newcommand{\ceta}{c_\eta}
\newcommand{\setatilde}{\widetilde{s}_\eta}
\newcommand{\cetatilde}{\widetilde{c}_\eta}
\newcommand{\tr}[1]{\mathrm{tr}\left[ #1 \right]}
\newcommand{\axial}{a}
\newcommand{\transpose}[1]{#1^\mathrm{t}}
\newcommand{\vevsh}{\sh^*}
\newcommand{\vevsetatilde}{\widetilde{s}_\eta^*}
\newcommand{\vevcetatilde}{\widetilde{c}_\eta^*}
\newcommand{\gb}[1]{{\hat #1}}

\addtokomafont{disposition}{\rmfamily\boldmath}
\addtokomafont{caption}{\small}
\addtokomafont{captionlabel}{\small\bfseries}

\title{\Large Electroweak symmetry breaking and collider signatures in the next-to-minimal
composite Higgs model}%
\author{\large
Christoph Niehoff,
Peter Stangl,
and David M. Straub%
\footnote{E-Mail:
\texttt{\href{mailto:christoph.niehoff@tum.de}{christoph.niehoff@tum.de},
\href{mailto:peter.stangl@tum.de}{peter.stangl@tum.de},
\href{mailto:david.straub@tum.de}{david.straub@tum.de}}}
}

\date{\normalsize\itshape
Excellence Cluster Universe, TUM, Boltzmannstr.~2, 85748~Garching,
Germany}

\begin{document}

\maketitle

\bigskip

\begin{abstract}
\noindent
We conduct a detailed numerical analysis of the composite pseudo-Nambu-Goldstone
Higgs model based on the next-to-minimal coset $\text{SO}(6)/\text{SO}(5)\cong\text{SU}(4)/\text{Sp}(4)$,
featuring an additional SM singlet scalar in the spectrum, which we allow to mix with the Higgs boson.
We identify regions in parameter space compatible with all current experimental
constraints, including radiative electroweak symmetry breaking,
flavour physics, and direct searches at colliders.
We find the additional scalar, with a mass predicted to be below a TeV,
to be virtually unconstrained by current LHC data, but potentially
in reach of run 2 searches.
Promising indirect searches include rare semi-leptonic $B$ decays,
$C\! P$ violation in $B_s$ mixing,
and the electric dipole moment of the neutron.
\end{abstract}

\setcounter{tocdepth}{2}
\tableofcontents

\section{Introduction}

The idea that the Higgs is a composite object is an appealing road to address the hierarchy problem of the Standard Model (SM) of particle physics.
It necessitates the existence of a new, strongly interacting sector besides the known SM particles, that introduces heavy composite resonances which can possibly show up in direct and indirect searches for New Physics (NP).
Introducing the idea of partial compositeness~\cite{Kaplan:1991dc}, severe constraints from flavour physics can be circumvented and a connection to extra-dimensional theories can be drawn.
Furthermore, the lightness of the Higgs boson can be explained if the new strong sector is endowed with a global symmetry $G$ that is spontaneously broken to $H$, forming a coset $G/H$ under which the Higgs is a (pseudo-)Nambu-Goldstone boson (pNGB).
If the theory is subject to a small explicit breaking of this global symmetry, then the Higgs obtains a mass that is naturally suppressed compared to the scale of New Physics~\cite{Kaplan:1983fs, Dugan:1984hq}.

Even without specifying the physics of the strongly interacting sector,
the low-energy physics of this pNGB Higgs can be described by an
effective theory that depends mainly on the symmetry breaking structure, i.e.\ the coset $G/H$.
A minimal requirement on this coset is that the global symmetries include a custodial symmetry $\text{SU}(2)_\text{L} \times \text{SU}(2)_\text{R} \cong \text{SO}(4)$ that protects the electroweak precision observable $T$ from large deviations~\cite{Agashe:2004rs}.
The minimal coset satisfying this is $\text{SO}(5)/\text{SO}(4)$,
leading to four NGB degrees of freedom that can be identified with the Higgs doublet.
This minimal coset has been studied extensively in the literature starting from~\cite{Agashe:2004rs},
and scrutinized in a thorough numerical analysis by us recently~\cite{Niehoff:2015iaa}.
In this work we extend this analysis to the next-to-minimal coset $\text{SO}(6)/\text{SO}(5)\cong\text{SU}(4)/\text{Sp}(4)$,
featuring one additional pNGB degree of freedom~\cite{Georgi:1984af,Gripaios:2009pe,Serra:2015xfa,Low:2015qep,Bellazzini:2015nxw,Redi:2012ha}, which we will denote $\eta$.
From a theoretical point of view, this coset construction is appealing as it is the minimal one that contains a Higgs doublet and can arise from a global symmetry broken by a fermion-bilinear condensate in a UV completion\footnote{For a less minimal construction employing a supersymmetric gauge theory that gives rise to an $\text{SO}(5)/\text{SO}(4)$ coset, see \cite{Caracciolo:2012je}.}~\cite{Katz:2005au,Galloway:2010bp,Barnard:2013zea,Ferretti:2013kya,Cacciapaglia:2014uja,Sannino:2016sfx,Galloway:2016fuo,Agugliaro:2016clv}.
Furthermore, it could give important contributions to baryogenesis in the early universe as the effective Higgs potential can give rise to a strongly first-order electroweak phase transition and new sources of $C\! P$ violation~\cite{Espinosa:2011eu}.
If the coset is supplemented by a suitable $\mathbb{Z}_2$ parity (such that the coset becomes $\text{O}(6)/\text{O}(5)$),
the additional scalar is stable and could serve as a dark matter candidate~\cite{Frigerio:2012uc}.
In this work, we will instead consider the case where $\eta$ can decay to SM states, resulting in interesting collider signatures.
In general, the otherwise pseudoscalar state $\eta$ can also mix
with the Higgs, leading to a spontaneous breaking of $C\! P$ in the scalar sector
and to a modification of Higgs physics.

For a special choice of couplings between right-handed elementary quarks and the composite sector, the pseudoscalar $\eta$ does not get a vacuum expectation value and effectively decouples from the SM fields.
In this limit, the model essentially behaves very similar to the minimal $\text{SO}(5)/\text{SO}(4)$ case.

The goal of this work is the following.
Similar to the analysis in~\cite{Niehoff:2015iaa} we want to analyse the non-minimal coset in an extensive numerical scan.
Due to the large dimensionality of the parameter space we are forced to use Markov Chain techniques for sampling the regions of parameter space that are compatible with the current experimental data.
We take into account a large range of constraints coming from indirect probes such as electroweak precision observables, modifications of $Z$ boson couplings, Higgs physics and quark flavour observables as well as direct searches at colliders.
We pay special attention to generating a scalar potential that gives rise to realistic electroweak symmetry breaking and is calculable at the one-loop level.
For a concrete realization of the non-mininal coset, we have adopted the ``4D Composite Higgs Model''~\cite {DeCurtis:2011yx}, modified to the $\text{SO}(6)/\text{SO}(5)$ coset.
To avoid stringent flavour constraints,
we further supplement it with a $\text{U}(2)^3$ flavour symmetry,
that was already shown to successfully reconcile flavour and electroweak precision constraints
with a relatively light compositeness scale in the case of the minimal
coset~\cite{Barbieri:2012tu,Niehoff:2015iaa}.

Another important feature of this non-minimal coset is that Wess-Zumino-Witten (WZW) terms~\cite{Wess:1971yu,Witten:1983tw} are admitted~\cite{Gripaios:2009pe}.
They give rise to additional $\eta Z Z$, $\eta W^+ W^-$ and $\eta G G$ couplings and can therefore influence the constraints from direct searches in these channels~\cite{Gripaios:2009pe,Bellazzini:2015nxw,Low:2015qep}.
In the absence of a specific UV completion, we choose however to
neglect possible contributions from the WZW terms in this work
as they would only make the constraints stronger; moreover,
being treated as free parameters the WZW could simply be tuned
to zero by our scan to avoid additional constraints.

One should keep in mind that CHMs in general are non-renormalizable and, therefore, they could be strongly affected by UV physics.
The cutoff of these theories is usually given as $4 \pi f=\mathcal{O}(10 \, \text{TeV})$, such that considerable theoretical uncertainties in the calculations are to be expected.
Nevertheless, we restrict ourselves to observables that are not strongly sensitive to UV contributions and are thus able to apply the experimental bounds in a reasonable way.

The outline of this work is as follows.
In section \ref{sec:model}, we review the model that we implemented for this work in detail, placing special emphasis on the structure of the Higgs potential.
Our numerical approach as well as the constraints used are layed out in section \ref{sec:Analysis}.
We present our results in section \ref{sec:results}, where the most significant phenomenology is that of the new scalar $\eta$.
We conclude in section \ref{sec:conclusion}.

\section{Model} \label{sec:model}
In this section, we discuss how to realize a concrete model that is phenomenologically viable and contains a Higgs doublet emerging from an $\text{SO}(6)/\text{SO}(5)$ coset.
The main features of this coset were already elaborated in~\cite{Gripaios:2009pe} in the language of $\text{SU}(4) / \text{Sp}(4)$.

To be able to study radiative electroweak symmetry breaking in the low-energy
effective theory and determine whether the correct SM vacuum can be reproduced,
the model has to be constructed such that the scalar potential is
calculable, i.e.\ finite, at least at the one-loop level.
One possibility to guarantee a calculable potential is to use a model that is deconstructed from an extra-dimensional gauge theory.
Therefore, we adapt the minimal construction of the ``4D Composite Higgs''~\cite {DeCurtis:2011yx} with two sites and modify it to the $\text{SO}(6)/\text{SO}(5)$ case in a straight-forward way.\footnote{
Alternative possibilities would be the ``Discrete Composite Higgs model'' \cite{Panico:2011pw}, where one-loop calculability of the effective potential is given only for a three-site construction, or ``General Composite Higgs Models'' \cite{Marzocca:2012zn}, for which one has to enforce Weinberg Sum Rules explicitly.}
This means there will be elementary states and only one level of composite resonances; for further details on the construction we refer to the original publication~\cite {DeCurtis:2011yx}.

\subsection{Boson sector}\label{sec:BosonSector}
The global symmetry breaking pattern is parametrized as usual by the NGB matrix
\begin{equation}
 \U = \exp \left[ \im \frac{\sqrt{2}}{f} \, \Pi(x) \right],
\end{equation}
where $\Pi(x) = \sum \limits_{i=1}^4 \pi_i(x) \T^i_{\hat 2} + \pi_5(x) \T_S$ denotes the NGB degrees of freedom with the generators $\T^i_{\hat 2}$ and $\T_S$ given in appendix \ref{appendix:SO6_generators} and $f$ is the symmetry breaking scale.
Out of these 5 NGBs, 3 are eaten by the SM gauge bosons, leaving two physical states, $\pi_4$ and $\pi_5$.
It is convenient to parametrize these two fields by modulus and angle in the $\pi_4-\pi_5$ plane~\cite{Redi:2012ha},
\begin{equation}
 \pi_4 = \gb{h} \cos\left( \frac{\gb{\eta}}{f} \right), \qquad \pi_5 = \gb{h} \sin\left( \frac{\gb{\eta}}{f} \right),
\end{equation}
giving the scalar $\gb{h}$ and a SM-singlet pseudoscalar $\gb{\eta}$.\footnote{
At this point we want to introduce the following notation: with $\gb{h}, \gb{\eta}$ we denote both scalar fields in the gauge basis.
Through the effective potential, both fields can obtain a vacuum expectation value and possibly mixing terms between them are introduced.
Because of these mixings, the physical Higgs field $h$ and the addition scalar $\eta$ in the mass basis are linear combinations of the gauge basis states.
In the above construction, $\gb{h}$ and $\gb{\eta}$ are $C\! P$ even and odd, respectively.
Consequently, the mass basis states do not have a defined $C\! P$ parity, but rather they have scalar and pseudoscalar admixtures depending on the scalar mixing angle $\alpha$ (cf. discussion in \ref{sec:ScalarPotential}).
}
The NGB matrix then takes the form
\begin{equation}
 \mathcal{U} =
 \left( \begin{array}{cccccc}
 1 &  &  &                     &                      &          \\
  & 1 &  &                     &                      &          \\
  &  & 1 &                     &                      &          \\
  &  &  &  \ch \ceta^2+\seta^2 &  -(1-\ch) \seta \ceta & \sh \ceta \\
  &  &  & -(1-\ch) \seta \ceta & \ch \seta^2 + \ceta^2 & \sh \seta \\
  &  &  &           -\sh \ceta &            -\sh \seta &       \ch
\end{array} \right), \label{eq:GoldstoneMatrix}
\end{equation}
where we introduced the usual notations $\sh = \sin(\gb{h} / f)$, $\seta = \sin(\gb{\eta} / f)$, $\ch = \cos(\gb{h} / f) = \sqrt{1-\sh^2}$ and $\ceta = \cos(\gb{\eta} / f) = \sqrt{1-\seta^2}$.

In the minimal 4DCHM model~\cite {DeCurtis:2011yx}, the sector of electroweak composite vector bosons is modeled using the two symmetry breaking cosets $(\text{SO}(6)^1_\text{L}~\times~\text{SO}(6)^1_\text{R}) / \text{SO}(6)^1_{\text{L}+\text{R}}$ and $\text{SO}(6)^0 / \text{SO}(5)$.
By gauging the $\text{SU}(2)_\text{L}~\times~\text{U}(1)$ subgroup of $\text{SO}(6)^1_\text{L}$ one introduces elementary gauge bosons similar to the SM ones. The heavy composite vectors are parametrized as gauge bosons of the gauged diagonal group of $\text{SO}(6)^1_\text{R}$ and $\text{SO}(6)^0$.
Using these fields, the gauge Lagrangian takes the simple form
\begin{equation}
 \mathcal{L}_\text{gauge} = - \frac{1}{4} \tr{A_{\text{elem} \, \mu\nu} A_\text{elem}^{\mu \nu}} \,\,-\,\, \frac{1}{4} \tr{\rho_{\mu \nu} \rho^{\mu \nu}},
\end{equation}
where the resonances are in the 15-dimensional adjoint representation of $\text{SO}(6)$,
\begin{equation}
 \rho^\mu = \rho^\mu_a \T^{a} = \rho_L^\mu + \rho_R^\mu + \axial_1^\mu + \axial_2^\mu + \rho_S^\mu,
\end{equation}
with two triplets $\rho_L^\mu \in (\mathbf{3},\mathbf{1})$, $\rho_R^\mu \in (\mathbf{1},\mathbf{3})$, two bidoublets $\axial_1^\mu, \axial_2^\mu \in (\mathbf{2},\mathbf{2})$ and one singlet $\rho_S^\mu \in (\mathbf{1}, \mathbf{1})$ under $\text{SU}(2)_\text{L} \times \text{SU}(2)_\text{R}$.
Only the components $\axial_2^\mu$ and  $\rho_S^\mu$ correspond to the $\text{SO}(6)/\text{SO}(5)$ coset directions.

In this construction, the $\sigma$-model fields couple to the vector bosons through covariant derivatives,
\begin{equation} \label{eq:BosonLagrangian}
 \mathcal{L}_\sigma = \frac{f_1^2}{4} \tr{\left( \mathcal{D}_\mu \Omega_1  \right)^\dagger \left( \mathcal{D}^\mu \Omega_1  \right)} + \frac{f_2^2}{2} \left[ \transpose{\left( \mathcal{D}_\mu \Omega_2 \right)} \left( \mathcal{D}^\mu \Omega_2 \right) \right]_{66},
\end{equation}
where
\begin{align}
 \mathcal{D}_\mu \Omega_1 &= \partial_\mu \Omega_1 - \im \left( g_0 W_{\mu}^a \T^{a}_L + g'_0 B_\mu \T^{3}_R \right) \Omega_1 + \im g_\rho \Omega_1 \rho_\mu, \\
 \mathcal{D}_\mu \Omega_2 &= \partial_\mu \Omega_2 - \im g_\rho \, \rho_\mu \, \Omega_2.
\end{align}
Here, $W_\mu$ and $B_\mu$ denote the elementary gauge fields with couplings $g_0$ and $g'_0$, respectively.

In the context of this deconstructed model, we adopt the so-called ``holographic gauge'' \cite{Panico:2007qd, Marzocca:2012zn}, where the pNGBs only appear through the mixings of elementary and composite states and not within the composite sector itself.
This means that effectively one sets $\Omega_1 \rightarrow \mathcal{U}$ and $\Omega_2 \rightarrow \mathds{1}$, where in $\Omega_1$ the symmetry breaking scale is identified with $f_1$.
For this choice, there are mixings of composite vector bosons with the pNGBs that have to be removed by field redefinitions of the vector resonances followed by rescalings of the pNGBs.
As a consequence, all dependences on the pNGBs will be through
\begin{equation} \label{eq:sh_setatilde_def}
 \sh = \sin \left( \frac{\gb{h}}{f} \right) \quad \text{and} \quad \setatilde = \sin \left( \frac{\gb{\eta}}{f \sin \left( \frac{v_h}{f} \right)} \right),
\end{equation}
where $v_h = \left< \gb{h} \right>$ is the numerical value of the $\gb{h}$-vev.
The explicit redefinitions are given in appendix \ref{appendix:BosonFieldshift}.

In this model, the SM Higgs vev is given as
\begin{equation} \label{eq:vSM}
 v_\text{SM} = f \, \vevsh \, \vevcetatilde,
\end{equation}
where $\vevcetatilde = \sqrt{1-{\vevsetatilde}\,\!^2}$ and $\vevsh,\vevsetatilde$ denotes the minimum of the effective scalar potential (cf. sec. \ref{sec:ScalarPotential}).

To have a phenomenologically viable model, one has to enlarge the occurring symmetries.
In particular, this applies to QCD and hypercharge.
To accommodate the strong interaction, one expects a QCD-like symmetry in the composite sector with heavy gluon partners $\mathcal{G}_\mu$ coupling only to composite quark partners.
Furthermore, as usual in composite Higgs models with custodial protection, one has to introduce an additional $\text{U}(1)_X$ symmetry, such that hypercharge is given by
\begin{equation}
 Y = \T{}^3_R + X.
\end{equation}
Also for this symmetry, one then expects heavy resonances $X_\mu$.
In general, these additional resonances can mix with elementary gauge bosons such that we parametrize the mixing Lagrangian as follows:
\begin{equation}
 \mathcal{L} \supset \frac{f_X^2}{4} \left( g'_0 B_\mu - g_X X_\mu \right)^2 \,\, + \,\, \frac{f_G^2}{4} \left( g_{s0} G_\mu - g_{\rho 3} \mathcal{G}_\mu \right)^2.
\end{equation}

It is the key property of partial compositeness that the elementary states mix with the composite ones such that in the physical (mass) basis there are light eigenstates that are identified with the SM degrees of freedom and heavy states that we will call heavy resonances.
These mixings are given by non-diagonal mixing matrices which have to be diagonalized in order to get to the mass basis.
The mass mixing matrices for the vector bosons are shown in appendix \ref{appendix:MassMatrices:Vector}.

\subsection{Fermion sector} \label{sec:FermionSector}
In the fermion sector, one generally has the freedom to choose a representation of the global symmetry, in our case $\text{SO}(6)$, for the composite vector-like resonances.
Assuming a custodial protection of the $Z$ couplings~\cite{Agashe:2006at}, the simplest case is given by choosing the fundamental~$\mathbf{6}$.\footnote{
A model using the symmetric, traceless representation $\mathbf{20'}$ is presented in \cite{Serra:2015xfa}.
}
Under the custodial symmetry $\text{SU}(2)_\text{L} \times \text{SU}(2)_\text{R}$, the $\mathbf{6}$ decomposes into a bidoublet $Q$ and two singlets $S_1$, $S_2$,
\begin{equation} \label{eq:fundamental}
 \Psi_\mathbf{6} = \begin{pmatrix}
                     \begin{pmatrix}
			Q^{++} & Q^{+-} \\
			Q^{-+} & Q^{--}
		     \end{pmatrix} \\
			S_1 \\
			S_2
		  \end{pmatrix}.
\end{equation}

In the construction of~\cite {DeCurtis:2011yx}, the left-handed elementary fermions mix with the composite fundamental $\Psi$ while the right-handed ones mix with a different composite fundamental $\widetilde{\Psi}$.
Therefore, for each SM fermion there are \emph{two} composite resonances.
Going to the holographic gauge, the composite quark Lagrangian can be written as
\begin{align}
 \mathcal{L}_\text{quark} &= \overline \Psi_u \left( \im \slashed{\mathcal{D}} - m_U \right) \Psi_u + \overline{\widetilde{\Psi}}_u \left( \im \slashed{\mathcal{D}} - m_{\widetilde{U}} \right) \widetilde{\Psi}_u \nonumber \\
			  &- m_{Y_u} \left( \overline{Q}_{u \text{L}} \widetilde{Q}_{u\text{R}} + \overline{S}_{1\, u \text{L}} \widetilde{S}_{1 \, u \text{R}} \right) - (m_{Y_u} + Y_u) \, \overline{S}_{2\, u \text{L}} \widetilde{S}_{2 \, u \text{R}} \label{eq:compositeLagrangian} \\
			  & + (u \leftrightarrow d) \,\,\, + \,\,\, \text{h.c.} \nonumber
\end{align}

The mixings of the composite states with the elementary sector break the global symmetries explicitly.
This is parametrized by embedding the elementary fields into incomplete representations of $\text{SO}(6)$.
For the left-handed fields, it is clear how to embed them into the bidoublet components of the fundamental. But there is an ambiguity for the right-handed fields as in principle they could be embedded into each of the singlets in (\ref{eq:fundamental}).
So in general, they have to be embedded into a linear combination of both possibilities.
Then, one can write down the mixing terms
\begin{align}
 \mathcal{L}_\text{mix} &= \Delta_{u \text{L}} \, \overline{\xi}_{u \text{L}} \U \Psi_{u \text{R}} \nonumber \\
                        &+ \Delta^5_{u \text{R}} \, \overline{\xi}^5_{u\text{R}} \U \widetilde{\Psi}_{u\text{L}} \,\, + \,\, \Delta^6_{u \text{R}} \, \overline{\xi}^6_{u\text{R}} \U \widetilde{\Psi}_{u\text{L}} \label{eq:CompElemMixings} \\
                        & + (u \leftrightarrow d) \,\,\, + \,\,\, \text{h.c.}, \nonumber
\end{align}
where the elementary embeddings are given in appendix \ref{appendix:SO6_embeddings}.

In the following, to avoid excessive effects in flavour-changing neutral currents, we will always assume that the composite sector respects a flavour symmetry that is only broken via the mixings with the elementary sector.
Hence, all the parameters in the above composite Lagrangian (\ref{eq:compositeLagrangian}) have to be understood as diagonal matrices in flavour space, while the composite-elementary mixings (\ref{eq:CompElemMixings}) can have a flavour-violating structure that in the end has to reproduce the known CKM structure.
For doing this, it is enough that only the mixings of one chirality with the composite sector are actually breaking the flavour symmetry~\cite{Cacciapaglia:2007fw, Redi:2011zi, Barbieri:2012uh},
leading to CKM-like flavour violation and thus a maximal protection of flavour-changing neutral currents.
Since models based on a $\text{U}(3)^3$ flavour symmetry are plagued by problems
with electroweak precision tests and compositeness searches~\cite{Barbieri:2012tu},
we employ a $\text{U}(2)^3$ symmetry
acting on the first two generations \cite{Barbieri:2012uh}.
A major difference to the model with a minimal coset is the presence of
two composite-elementary mixings for the right-handed quarks,
cf.\ \eqref{eq:CompElemMixings}.
Thus, if only the mixings of left-handed quarks were flavour symmetric,
there would be two flavour violating structures for up-type quarks and
two for down-type quarks, leading to non-CKM-like flavour violation unless
some of the stuctures are assumed \textit{ad hoc} to be aligned. Thus,
we restrict ourselves to the case of breaking the flavour symmetries only through the mixings of \emph{left-handed} elementary fermions with the composite sector.
This scenario we will refer to as $\text{U}(2)^3$-\emph{right compositeness} ($\text{U}(2)^3_\text{RC}$).
The explicit form of the composite-elementary mixing is shown in appendix \ref{appendix:FlavourStructure}.

Just as for vector bosons, the elementary fermions mix with the composite ones due to partial compositeness.
The lightest states are the SM quarks, while all other states are heavy resonances.
In appendix \ref{appendix:MassMatrices:Fermion}, we give the mass mixing matrices in the quark sector.

In this work, we are mainly interested in the phenomenology of the quark sector.
For a full model, partial compositeness also has to be implemented for leptons, leading to heavy resonances with lepton quantum numbers.
Although in principle composite leptons can lead to an interesting phenomenology of the effective potential \cite{Carmona:2014iwa} and of flavour observables \cite{Niehoff:2015bfa}, they also lead to major complications for model building since one has to introduce some mechanism that generates the correct PMNS matrix and neutrino mixings.
Because of this, we neglect these effects and model the lepton sector in a trivial way as being purely elementary.
We explicitly include Yukawa couplings to the $C\! P$ even scalar $\gb{h}$ by taking the corresponding SM-value, but we do not consider couplings to the pseudoscalar $\gb{\eta}$.
As a consequence of this, a coupling of the SM leptons to $\eta$ in the mass basis is introduced only by its mixing with the Higgs, uniquely fixed by the scalar mixing angle $\alpha$.
It should be kept in mind that in a more complete model
these couplings could be different.

\subsection{Scalar potential} \label{sec:ScalarPotential}
As usual in Composite Higgs Models, the interactions with the elementary sector break the global symmetries of the composite sector explicitly, such that the Nambu-Goldstone-Higgs turns into a \emph{pseudo}-Nambu-Goldstone-Higgs and its mass can be explained.
This means that due to the composite-elementary mixings a Coleman-Weinberg potential with non-trivial minimum is generated.

The one-loop effective potential can be calculated by the Coleman-Weinberg formula~\cite{Coleman:1973jx},
\begin{align} \label{eq:EffPotGeneral}
 V_\mathrm{eff}(\gb{h},\gb{\eta}) &= \sum \frac{c_i}{64 \pi^2} \left( 2 \, \tr{M_i^2} \, \Lambda^2 - \tr{M_i^4} \log \left[ \Lambda^2 \right] \,\,  + \,\, \tr{M_i^4 \log \left[ M_i^2 } \right] \right),
\end{align}
where the sum goes over all particle species (except the pNGB states) and where ${M_i \equiv M_i(\gb{h},\gb{\eta})}$ denotes the $\gb{h}$- and $\gb{\eta}$-dependent mass matrices of the particular particle species (which are given in appendix \ref{appendix:MassMatrices}).
The constants $c_i$ depend on the spin and charge of the particles and take the values
\begin{equation}
 c_i = \left\{ \begin{array}{rl} 3 & \text{for neutral gauge bosons,} \\
                                 6 & \text{for charged gauge bosons,} \\
                               -12 & \text{for (coloured) Dirac fermions.} \\ \end{array} \right.  \nonumber
\end{equation}
As the M4DCHM~\cite {DeCurtis:2011yx} is deconstructed from an extra-dimensional theory, it is ensured that the so-called Weinberg sum rules are satisfied~\cite{Marzocca:2012zn},
\begin{align}
  &\tr{M_i^2} - \tr{M_i^2(\gb{h}=0,\gb{\eta}=0)}=0, \\
  &\tr{M_i^4} - \tr{M_i^4(\gb{h}=0,\gb{\eta}=0)} =0,
\end{align}
such that the potential is not UV-sensitive and thus calculable.%
\footnote{
 One has to keep in mind that the EFT description in this form is non-renormalizable.
 Therefore, calculability only refers to the effective potential at one-loop level and potentially higher-order corrections could change the picture significantly.
 We however do not consider this possibility.
}
In this case the potential simplifies to
\begin{equation}
 V_\text{eff}(\gb{h},\gb{\eta}) = \sum \frac{c_i}{64 \pi^2} m_i^4(\gb{h},\gb{\eta}) \, \log \left( m_i^2(\gb{h},\gb{\eta}) \right),
\end{equation}
where $m_i(\gb{h},\gb{\eta})$ denote just the $\gb{h}$- and $\gb{\eta}$-dependent masses of all particles in the mass basis.

From (\ref{eq:GoldstoneMatrix}), one sees that all dependence of $\gb{h}$ and $\gb{\eta}$ appears through the trigonometric functions $\sh$ and $\setatilde$, which are defined in~(\ref{eq:sh_setatilde_def}). So also the effective potential
\begin{equation}
 V_\text{eff}(\gb{h},\gb{\eta}) \equiv V_\text{eff}(\sh, \setatilde),
\end{equation}
only depends on them and we recall that their values at its minimum are denoted by $\vevsh$ and $\vevsetatilde$.
The scalar mass matrix is given by the Hessian of the potential at the minimum,
\begin{equation}
M^2_\text{scalar} = \left. \left( \begin{array}{cc}
                            \partial_\gb{h}^2 & \partial_\gb{h} \partial_\gb{\eta} \\ \partial_\gb{h} \partial_\gb{\eta} & \partial_\gb{\eta}^2
                           \end{array} \right) \,\, V_\text{eff}(\sh, \setatilde) \right|_{\sh=\vevsh, \setatilde=\vevsetatilde},
\label{eq:ScalarMassMatrix}
\end{equation}
whose eigenvalues correspond to the masses of the scalar states and whose off-diagonal elements measure the amount of scalar mixing and thus violation of $C\! P$.
This mass matrix is diagonalized by a $2 \times 2$ orthogonal matrix parametrized by the scalar mixing angle~$\alpha$.
In principle, $\eta$ could be lighter than the Higgs-like scalar $h$.
We however always identify the SM Higgs with the lightest scalar particle and therefore neglect the possibility of a light $\eta$.

The structure of the scalar potential was already thoroughly investigated in~\cite{Gripaios:2009pe}.
Let us summarize here their results and translate them into the $\text{SO}(6)/\text{SO}(5)$ language as used in this work.
There are two sources of explicit breaking of the remaining global $\text{SO}(5)$ symmetry.
By gauging the SM subgroup, gauge contributions are induced that lead to an effective potential for the Higgs.
However, this gauging only breaks $\text{SO}(5)$ down to $\text{SU}(2)_L \times \text{U}(1)_Y \times \text{U}(1)_S$, where the latter $\text{U}(1)$ is the symmetry generated by $\T{}_S$ under which $\pi_5$ shifts.
This means that, considering gauge contributions alone, $\pi_5$ remains a true massless NGB and thus no potential is generated for it.
As a consequence, the effective potential is only a function of $\pi_4 = \gb{h} \, \cos(\gb{\eta}/f)$, such that the minimum is realized for $\vevcetatilde=0$, i.e. $\vevsetatilde =1$.
In this case, one also finds from (\ref{eq:ScalarMassMatrix}) that the scalar sector necessarily contains a massless mode.

In order to prevent the $\eta$ from becoming massless, one has to break the $\text{U}(1)_S$ symmetry explicitly.
As the composite sector is invariant under it by construction, this symmetry can only be broken if the elementary fermion embeddings have a non-consistent $\text{U}(1)_S$ charge assignment.
The left-handed embeddings always respect this symmetry, but for the right-handed embeddings this is only the case if they are eigenvectors of $\T{}_S$, which corresponds to the choice
\begin{equation} \label{eq:ConsistentU1S}
 \Delta_\text{R}^5=\pm \im \Delta_\text{R}^6.
\end{equation}
For other choices of relative phases (i.e. the $\phi^6$-parameters in (\ref{eq:FlavourSpurions})) and absolute values, $\eta$ is a massive pNGB.
Considering a $\text{U}(2)^3_\text{RC}$ flavour structure\footnote{For flavour structures with left-compositeness, the right-handed composite elementary mixings are necessarily off-diagonal. As a consequence, $\eta$ will always be massive for that case.}, there are four different $\Delta^{5,6}_\text{R}$-coefficients that can break the $\text{U}(1)_S$ symmetry\footnote{For the up- as well as down-sector, there are separate coefficients for the first two and the third generation, cf. (\ref{eq:FlavourSpurions})}.
Depending on the source of $\text{U}(1)_S$ breaking, the mass of $\eta$ will take different values.
If this is done via the composite-elementary mixing of the top, then one expects $m_\eta$ to be naturally of the order of $f = \mathcal{O}(500-1000 \, \text{GeV})$.
However, in case $\text{U}(1)_S$ is respected in the top sector its breaking is less severe and one expects a much lighter $m_\eta$~\cite{Frigerio:2012uc}.

An interesting special case is given in the limit where all $\Delta^5_\text{R} \rightarrow 0$.
In this case the pseudoscalar $\gb{\eta}$ takes the trivial vev at $\vevsetatilde=0$ which does not break $C\! P$.
Hence, there is no mixing between $\gb{h}$ and $\gb{\eta}$ and thus $h=\gb{h}$ and $\eta=\gb{\eta}$.
Furthermore, also the couplings of $\eta$ to the SM fermions vanish, such that effectively $\eta$ decouples from the theory.
For this choice of parameters the relations (\ref{eq:ConsistentU1S}) cannot be satisfied if one wants to generate Yukawa couplings between the SM fermions and the Higgs, such that $\text{U}(1)_S$ is always explicitly broken and $\eta$ obtains a non-vanishing mass.
In practice, the model in this case looks similar to the M4DCHM with coset $\text{SO}(5)/\text{SO}(4)$.
As one can see from the mass matrices given in appendix \ref{appendix:MassMatrices}, the additional degrees of freedom (as compared to the pure $\text{SO}(5)/\text{SO}(4)$ model) also decouple for vanishing $\vevsetatilde$.

\section{Strategy} \label{sec:Analysis}
The aim of this work is to analyze the non-minimal composite Higgs introduced above numerically by sampling the parameter space in regions that are compatible with all experimental constraints.
For doing this, we adapt the numerical procedure of~\cite{Niehoff:2015iaa} which was used to analyze the parameter space of the minimal 4DCHM based on the coset $\text{SO}(5)/\text{SO}(4)$, and extend it to the model considered in this work.
In section \ref{sec:numerical_analysis} we will summarize the approach. For details on the scanning procedure, we refer to~\cite[Sec.~4.1]{Niehoff:2015iaa}.
All experimental constraints that are used in the numerical analysis are listed in section~\ref{sec:constraints}.

\subsection{Numerical Analysis}\label{sec:numerical_analysis}

For each parameter point $\vec \theta$, we calculate the one-loop effective potential and determine its minimum in the $\gb{h}$- and $\gb{\eta}$-direction,
allowing us to calculate the vev's in the scalar sector as well as the masses and couplings of all states.
We parametrize the experimental constraints by a $\chi^2$ function,
\begin{equation}
 \chi^2(\vec \theta) = \sum \limits_{i,j} \left( \mathcal{O}^\text{theo}_i(\vec \theta) - \mathcal{O}^\text{exp}_i \right) \left[ C^{-1} \right]_{i j}  \left( \mathcal{O}^\text{theo}_j(\vec \theta) - \mathcal{O}^\text{exp}_j \right),
 \label{eq:chi2}
\end{equation}
where $C$ is the covariance matrix containing experimental and theoretical uncertainties and we also take into account correlations e.g. for electroweak precision observables or in meson mixing.
The challenge is then to find the regions in the space of parameters $\vec \theta$ where this $\chi^2$ function is sufficiently small.
For the model at hand, the parameter space is 52-dimensional and, due to partial compositeness,
even the masses and couplings of the SM particles
are complicated functions depending on many model parameters in a non-trivial way, such that sophisticated numerical techniques are required.
After generating a random starting point that fulfills very basic requirements (such as e.g. a non-vanishing vev for at least one scalar) we employ the optimization tool \texttt{NLopt} \cite{NLopt}
to burn in into a region of parameter space that is close to a minimum.
When we have found such a region, we use adaptive Markov Chains Monte Carlos (MCMC) using the package \texttt{pypmc} \cite{pypmc} to sample the good parameter region.
Due to the properties of Markov Chains, the points retained after burn-in
are all globally in agreement with all experimental constraints imposed.
In addition to this condition on the global $\chi^2$, we also discard points where any
\textit{individual} constraint is violated by more than $3 \, \sigma$.
We will call all points passing these criteria the ``viable parameter points''
in the following.
As we do not know the total number of independent minima in parameter space, we have to rely on a large number of chains sampling in many different parameter regions.
Thus, we conducted our numerical scans on the Computational Center for Particle and Astrophysics (C2PAP) located in Munich.
In the end, we found 125
chains that successfully sampled in regions of parameter space with a satisfactory value of $\chi^2$.
We stress that we are not aiming at a statistical analysis of parameter space which would require us to obtain sufficient coverage and study
the impact of our choice of priors. Rather, we use the MCMC as a tool to find as many viable parameter points as possible, which is
not feasible with a blind parameter scan due to partial compositeness.

A possible limitation of this procedure is that
parameter points featuring small values of $\vevsetatilde$ are hard to find
as this requires reaching hyper surfaces of the parameter space where certain relations are fulfilled.
Due to a volume effect in high dimensions,
the Markov Chain is not likely to sample this region well.
To overcome this problem and also generate parameters in the limit $\vevsetatilde \approx 0$
we performed dedicated scans starting from points where the relation $\Delta^5_\text{R} \ll \Delta^6_\text{R}$ is explicitly enforced for all $\Delta^{5,6}_\text{R}$.
In the end, about one third of all chains were started in this limit.

\subsection{Constraints}\label{sec:constraints}

In this section, we summarize the experimental constraints that contribute to the $\chi^2$ function (\ref{eq:chi2}).
Since most observables have already been discussed in our study of the $\text{SO}(5)/\text{SO}(4)$ model,
we refer to~\cite[Sec. 3]{Niehoff:2015iaa} for details;
Here we just recapitulate the constraints used and point out the differences to the earlier analysis.

\begin{itemize}
 \item The most immediate constraints come from reproducing the SM masses for quarks, leptons, gauge bosons and the Higgs.
       We calculate the Higgs mass from the one-loop scalar potential using (\ref{eq:ScalarMassMatrix}), where we do not restrict the mass of $\eta$.
       The masses of the quarks are calculated at tree level by diagonalizing their mass matrices and interpreting them as $\overline{\text{MS}}$ running masses at the scale $m_t$.
       For the light quarks, we use the QCD RG evolution to connect them to the experimental results from lower scales.
       EWSB is included by demanding that the minimum of the effective potential reproduces the correct Fermi constant, such that the Higgs-vev has the right value.
 \item Reproducing the correct CKM matrix is imposed by
       including the constraints on the absolute values of CKM elements and on the angle $\gamma$ from processes that occur at tree level in the SM.
       Due to partial compositeness, the CKM matrix, defined as a $3 \times 3$ submatrix of the total quark mixing matrix, is not unitary and thus the appearing deviations from unitarity have to be restricted to small values.
       However, since we consider the $\text{U}(2)^3_\text{RC}$ flavour structure, these
       deviations are typically small (cf.~\cite{Niehoff:2015iaa}).
 \item Generally, electroweak precision observables, such as the $S$- and $T$-parameter, are important constraints for CHMs.
       The model based on the coset $\text{SO}(6)/\text{SO}(5)$ possesses a custodial symmetry in the Higgs sector, prohibiting tree-level contributions to the $T$-parameter.
       At the one-loop level, we include the dominant fermion contributions.
       In contrast, the $S$-parameter can already appear at tree-level, acting effectively as a lower bound on the masses of spin-1 resonances.
       Experimentally, the allowed ranges for the $S$- and $T$-parameters are strongly correlated, such that a deviation in one parameter has to be accompanied by a deviation in the other one.
 \item Considering fundamental representations for the fermion resonances includes an effective custodial protection of $Z_{b_\text{L} \bar b_\text{L}}$ vertices~\cite{Agashe:2006at}.
       However, these observables still give important constraints on the compositeness of the SM particles \cite{Straub:2013zca}.
       In our analysis, we include tree-level contributions to the $Z$ width in the channels $Z \rightarrow b \bar b$, $Z \rightarrow c \bar c$ and $Z \rightarrow q \bar q$, where $q$ runs over all quarks but the top.
 \item To constrain non-linearities arising from the pNGB nature of the Higgs, we include Higgs signal strengths.
       We calculate the signal strength from gluon fusion production in the decay channels $h \rightarrow WW$, $h \rightarrow ZZ$ and $h \rightarrow \tau^+ \tau^-$ at tree-level while we include one-loop effects for $h \rightarrow \gamma \gamma$.
       Compared to the analysis in~\cite{Niehoff:2015iaa}, we updated the experimental input and used the data given in~\cite[Table 8]{Khachatryan:2016vau}, showing combined ATLAS and CMS results from LHC run 1 at $7 \, \text{TeV}$ and $8 \, \text{TeV}$.

       As in our implementation of the 4DCHM the lepton sector is considered as being completely elementary, the signal strength in the $\tau \tau$-channel is directly proportional to $\cos(\alpha)$ where $\alpha$ is the scalar mixing angle.
       Hence, this channel would be a very good way to constrain the mixing $\gb{h}$ and $\gb{\eta}$ and thus $C\! P$ violation in the scalar sector.
       Given the large experimental uncertainties for this channel at present, this bound is still weak.
 \item Strong constraints on models with partial compositeness arise from flavour physics.
       We include meson-antimeson mixing in the $K^0$-, $B_d$- and $B_s$-systems,
       calculating the tree-level contributions to the mass differences for all three cases, the mixing phases for $B_d$ and $B_s$ as well as $\epsilon_K$ parameter for indirect $C\! P$ violation in the $K^0$ system.
       The theory uncertainty in these calculations depend on bag parameters that have to be determined from lattice QCD.
       Compared to~\cite{Niehoff:2015iaa} we used the updated values from the FNAL/MILC collaborations~\cite{Bazavov:2016nty}
       with considerably reduced uncertainties.

       Furthermore, we include constraints from rare $B$-decays.
       For this we calculate the branching ratios of the processes $b \rightarrow s \, \gamma$ and $B_s \rightarrow \mu \mu$.

\item Due to the presence of spontaneous $C\! P$ violation in the scalar sector,
we also include the neutron electric dipole moment (EDM) as a constraint,
computing the one-loop contributions to the quark EDMs and chromo EDMs involving
both SM states, the $\eta$ scalar, and heavy resonances, and using the sum rule
expression~\cite{Pospelov:2000bw,Jung:2013hka}
\begin{equation}
d_n = \left(1 ^{+0.5} _{-0.7}\right) \left[ 1.4\left( d_d - \tfrac{1}{4} d_u\right) + 1.1 e \left( \tilde{d}_d + \tfrac{1}{2} \tilde{d}_u\right)  \right] .
\label{nEDM}
\end{equation}
In our numerical analysis, we take into account the theory uncertainty in a
conservative way by imposing the experimental bound on the sum rule prediction
with the low value $0.3$ for the prefactor in brackets.

We do not take into account constraints from the electron EDM. Since we assume leptons to
be elementary, contributions arise first at the two-loop level \cite{Espinosa:2011eu}
and we find them to be subleading.

\item The compositeness of first-generation quarks can be constrained by four-quark contact interactions that contribute to the dijet angular distribution at the LHC.
We update our analysis in \cite{Niehoff:2015iaa}, that was based on \cite{Domenech:2012ai}, by computing the relevant dependence on proton
PDFs for LHC with a center-of-mass energy of 13~TeV to be able to include constraints from LHC run 2.
\item The heavy fermion and vector boson resonances as well as the scalar $\eta$ decay to SM particles and can thus be searched for at the LHC.
We predict the production cross section and the branching ratios into all possible SM final states $i,j$ for each of these heavy resonances $R$.
To this end, we calculate all possible partial widths $\Gamma_{R\to ij}$ at leading order. While the branching ratios are then simply given by
$\text{BR}(R\to ij)={\Gamma_{R\to ij}}/{\sum_{k,l}\Gamma_{R\to kl}}$,
for the production cross sections of vector bosons and the scalar $\eta$, we use the narrow-width approximation (NWA), employing~\cite{Olive:2016xmw,Chivukula:2016hvp}
\begin{equation}\label{eq:xsec}
\sigma_{pp\to R}= \frac{16\,\pi^2\,S_R\,c_R}{m_R}\,\sum_{i,j} \frac{1+\delta_{ij}}{S_i\,S_j\,c_i\,c_j}\,\Gamma_{R\to ij}\,\frac{\mathcal{L}_{ij}(s,m_R)}{s},
\end{equation}
where $S$ and $c$ count the number of polarizations and colours of inital and final states, $m_R$ is the mass of the resonance, $s$ is the center of mass energy of the collider squared and $\mathcal{L}_{ij}(s,m_R)$ is the parton luminosity of partons\footnote{We include the vector-boson-fusion (VBF) process by means of the effective W approximation (EWA) \cite{Dawson:1984gx,Chanowitz:1984ne,Altarelli:1987ue,Pappadopulo:2014qza}} $i$ and $j$ in a proton-proton collision with collider energy $\sqrt{s}$ and center of mass energy of partons $\sqrt{\hat{s}}=m_R$. In the gluon fusion production of $\eta$, we include a K-Factor of $2$ to approximate higher order corrections.
For the production cross section of heavy quarks we consider the model-independent results for pair-production obtained with the \texttt{Hathor} package~\cite{Aliev:2010zk}.

Our predictions are compared to experimental searches for bosonic resonances (cf. tables \ref{tab:exp_scalar_res} and \ref{tab:exp_vector_res} in Appendix~\ref{app:bosons}) and heavy vector-like quarks (cf. table \ref{tab:exp_quark_res} in Appendix~\ref{app:fermion}).
In addition to the experimental analyses included in~\cite{Niehoff:2015iaa}, we have implemented several new LHC searches for spin-1 and fermion resonances with $\sqrt{s}=13\,\text{TeV}$, as well as searches for spin-0 resonances with $\sqrt{s}=8\,\text{TeV}$ and $\sqrt{s}=13\,\text{TeV}$.

\end{itemize}

\section{Results} \label{sec:results}

\subsection{Scalar potential and fine-tuning}
\begin{figure}
 \centering
 \includegraphics[keepaspectratio=true,width=0.48\textwidth]{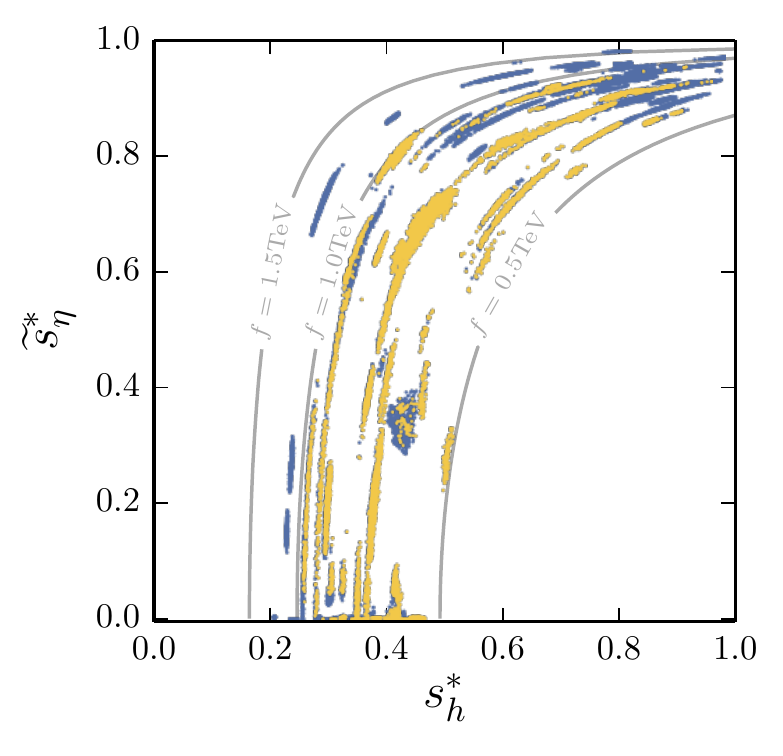}
\quad
 \includegraphics[keepaspectratio=true,width=0.48\textwidth]{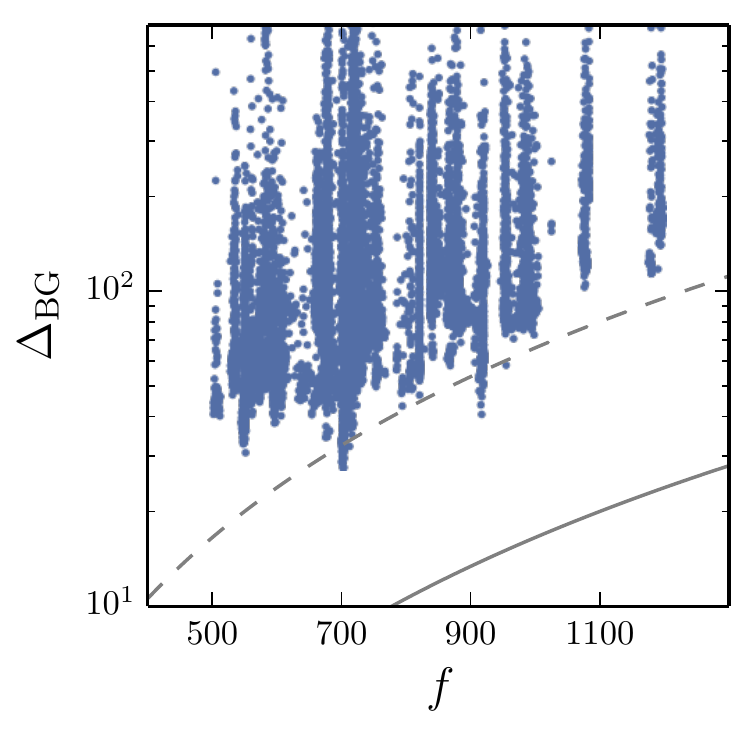}
 \caption{Left:
Location of the minima of $V_\text{eff}$ in the $(\vevsh$,$\vevsetatilde)$ plane found by the Markov chains.
In gray there are lines of fixed $f$ where $v_\text{SM} = 246\,\text{GeV}$ is realized.
In yellow we mark the points with a good fine-tuning ($\Delta_\text{BG}<100$).
Right:
Barbieri-Giudice measure $\Delta_\text{BG}$ of fine-tuning for ca. 40\% of the viable points and its correlation with $f$.
The gray lines show naive expectations for the finetuning (see main text).}
 \label{fig:sh_vs_setatilde}
\end{figure}

Electroweak symmetry breaking is characterized by the location $(\vevsh, \vevsetatilde)$ of the minimum of the effective scalar potential.
In principle, the minimum is allowed to take any value on the unit square $\left[ 0,1 \right] \times \left[ 0,1 \right]$, but demanding the correct SM Higgs vev $v_\text{SM} = 246 \, \text{GeV}$ (cf. eq. (\ref{eq:vSM})) restricts the allowed region for reasonable values of $f$.
In our scans we found values for $f$ roughly in the range $500 - 1200 \, \text{GeV}$.

In the left plot of figure~\ref{fig:sh_vs_setatilde}, we show predictions for the minimum of the effective potential, where we find viable parameter points for all reasonable values of $\vevsh$ and $\vevsetatilde$.
Generically, the scan has the tendency to yield parameter points for minima with large $\vevsh$ and $\vevsetatilde$.
However, points with $\vevsetatilde \approx 1$ are excluded as they imply a very small $m_\eta$ (see sec. \ref{sec:ScalarPotential}).
The region of $\vevsetatilde \approx 0$ is more tuned in the sense that it corresponds to the case of $\Delta^5_\text{R} \rightarrow 0$ which is hard to find in a general scan.
Therefore, we started dedicated scans with a preference for these values of parameters to have a good coverage of the small $\vevsetatilde$ region (cf. sec. \ref{sec:numerical_analysis}).
As remarked in sec. \ref{sec:ScalarPotential} this case is very similar to the M4DCHM with the minimal coset $\text{SO}(5)/\text{SO}(4)$.
One finds that for this case the found range of $\vevsh$ is comparable to the one found in~\cite{Niehoff:2015iaa}.
Only in the region of intermediate $\vevsetatilde$ the coverage is comparably low.

In the left plot of figure~\ref{fig:sh_vs_setatilde}, one can also identify the individual Markov chains as different clouds with constant $f$.
This is easy to understand from the fact that fermion and gauge boson contributions to the potential have to cancel each other to a rather large extend for guaranteeing the lightness of the Higgs~\cite{Marzocca:2012zn, Panico:2012uw}.
While the size of the gauge boson contributions is mainly driven by the parameter $f$, the fermion contributions depend on a large number of independent parameters, such as composite masses and composite-elementary mixings.
Thus, a change in $f$ would need a coordinated and collective change in many fermion parameters, which is again difficult to realize in a Markov Chain scan.

The amount of fine-tuning for the viable parameter points can be quantified using the Barbieri-Giudice measure~\cite{Barbieri:1987fn},
\begin{equation}
 \Delta_\text{BG} = \max \limits_{\lambda \in \text{parameters}} \left| \frac{\partial \log (m_Z)}{\partial \log (\lambda)} \right|,
\end{equation}
which gives the change in the electroweak scale when varying the fundamental parameters of the theory.
Unfortunately, the numerical evaluation of this measure is computationally expensive and thus we calculated the tuning for only about $40 \%$ of the viable parameter points.
They are shown in the right plot of figure~\ref{fig:sh_vs_setatilde}.
We find that a moderate tuning on the percent level is very well possible for $f < 1 \, \text{TeV}$ given the current experimental constraints.
For the least fine-tuned points we find a value of $\Delta_\text{BG} = 27$.
This is comparable to CHMs with a smaller coset~\cite{Niehoff:2015iaa, Panico:2012uw, Barnard:2015ryq}.

In figure \ref{fig:sh_vs_setatilde}, we further show the naive expectation for the minimal fine-tuning, ${\Delta_\text{min}\sim f^2/v_\text{SM}^2}$ as the solid gray line.
For the minimal CHM based on the coset $\text{SO}(5)/\text{SO}(4)$ with fermions in the fundamental representation it is well known that the potential is subject to a so-called double-tuning~\cite{Panico:2012uw}.
Using the notation of~\cite{Panico:2012uw}, it can be estimated by $\Delta_\text{min} \sim 1 / \epsilon^2 \times f^2/v_\text{SM}^2$ with $\epsilon<1$, i.e. the tuning is parametrically larger due to the particular structure of the potential.
We expect this to be true also for the next-to-minimal coset with fermions in the fundamental representation.
To guide the eye, we include the expectation for an exemplary value $\epsilon = 0.5$ as the dashed line in figure \ref{fig:sh_vs_setatilde}, which shows that our data points are consistent with double-tuning.

Further, in the left plot of figure~\ref{fig:sh_vs_setatilde} we also indicate the position of points with ${\Delta_\text{BG}<100}$ in the $\vevsh$-$\vevsetatilde$ plane.
This shows that a moderate fine-tuning can be archieved for all values of the vevs that allow for a not too large scale $f$.

\subsection{Higgs phenomenology}
\begin{figure}
 \centering
 \includegraphics[keepaspectratio=true,width=0.8\textwidth]{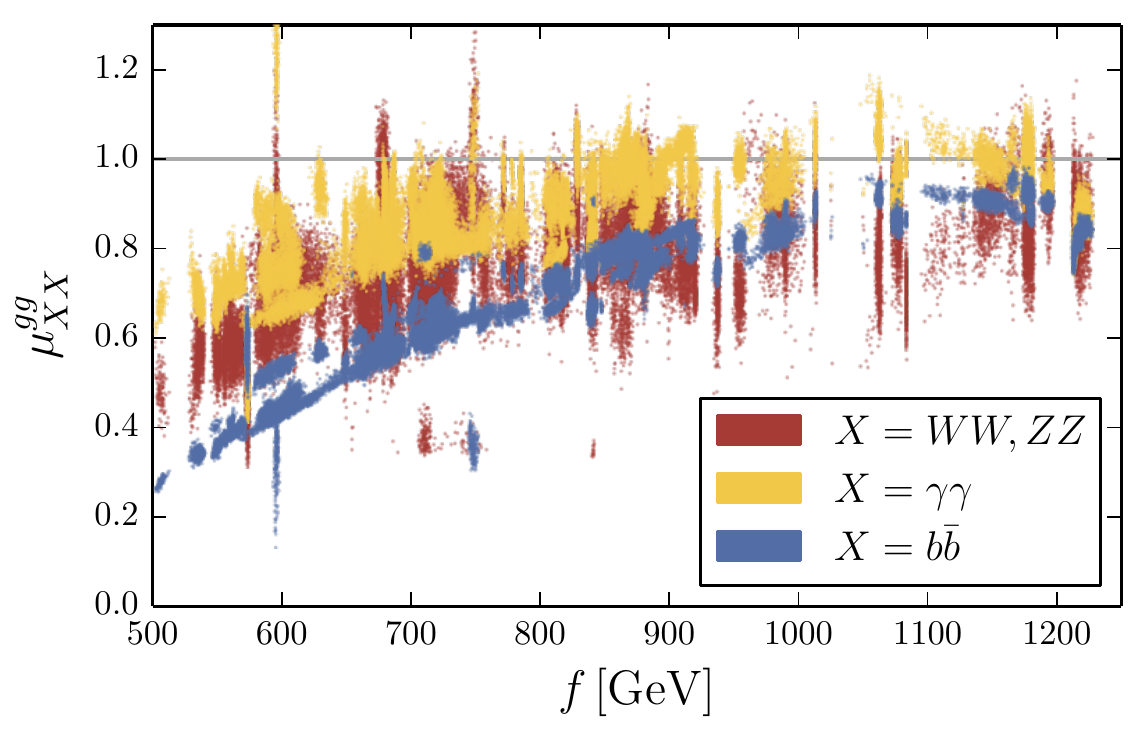}
 \caption{Modifications of Higgs signal strengths for different decay channels.
          The SM expectation is given by the gray line as $\mu^{gg}=1$.
          By custodial symmetry, the signal strengths into $WW$ and $ZZ$ are the same.}
 \label{fig:higgs_signalstrengths}
\end{figure}
A crucial feature of a CHM with a non-minimal coset is the enlarged scalar sector, such that modifications of the Higgs couplings can be induced by the mixing of $\gb{h}$ with the new pseudoscalar $\gb{\eta}$.
This mixing is induced by the effective potential via $\gb{h}$-$\gb{\eta}$ cross terms that strongly depend on whether $\gb{\eta}$ takes a vev.
In the case $\vevsetatilde=0$ there will be no mixing between the scalars and the Higgs couplings are only modified by the non-linear nature of the pNGB Higgs.
As soon as $\gb{\eta}$ takes a non-vanishing vev, the mixing sets in.

Limits on the modification of Higgs couplings can best be set in terms of Higgs signal strengths, which are defined for a certain channel $h \rightarrow X$ via the Higgs partial widths $r_X = \Gamma(h \rightarrow X) / \Gamma(h \rightarrow X)_\text{SM}$ as
\begin{equation}
 \mu^{gg}_X = \frac{r_X r_{gg}}{r_\text{tot}},
\end{equation}
where a production exclusively through $gg$-fusion is assumed and $r_\text{tot} = \Gamma_h/\Gamma_h^\text{SM}$.

At present, the experimental bounds on the Higgs signal strengths are still rather loose, allowing for even a vanishing signal strength at the few $\sigma$ level.
In figure~\ref{fig:higgs_signalstrengths} we show our results.
The overall dependence on the symmetry breaking scale $f$ is compatible with general considerations of Higgs coupling modifications in strongly-interacting theories~\cite{Giudice:2007fh} which predict $\mu^{gg}_X \sim (1 - c \frac{v_\text{SM}^2}{f^2})$, where $c$ is some model- and channel-dependent constant.
While these relations can be violated in the presence of a large degree of compositeness of light quarks~\cite{Delaunay:2013iia},
we find the deviations to be even much larger than in the case of the minimal
coset~\cite{Niehoff:2015iaa}. This is not suprising in
view of the mixing between the Higgs and $\eta$, which leads to an additional modification of the coupling between the SM particles and the Higgs.

\subsection{$\eta$ production and decay}
Being a pNGB, the scalar $\eta$ is usually\footnote{For 20\% of the viable parameter points, the masses of the lightest fermion resonances are slightly lighter than the $\eta$ mass.} the lightest state in the spectrum apart from the SM particles\footnote{While in principle $\eta$ could even be lighter than the Higgs, as noted in section \ref{sec:ScalarPotential}, we only discuss the case of $\eta$ being heavier.}.
With a total range of ca. $130-1600 \, \text{GeV}$, an interquartile range\footnote{The interquartile range is the range of values when cutting out the 25\% of points with the largest values and the 25\% of points with the smallest values.
While we use statistics vocabulary to describe the viable parameter points that we have found with our scanning procedure, we want to stress that we do not make statements about the propability of finding specific values. This is not possible due to the limitations of our scanning procedure discussed in secton \ref{sec:numerical_analysis}.} of ca. $550-790 \, \text{GeV}$ and a median of ca. $690 \, \text{GeV}$, the viable parameter points yield a mass for $\eta$ that lies well in the energy range accessible by the LHC.
However, we will see in this section that $\eta$ can escape all current direct bounds included in our scan.
In principle, the phenomenology of $\eta$ can also be constrained indirectly by observables sensitive to scalar four-fermion operators originating from a tree-level exchange of $\eta$.
In particular, these are contact interactings involving first-generation quarks and heavy meson-antimeson mixing.
Furthermore, $\eta$ could show up in penguin-induced flavour transitions such as $b \to s \gamma$.
For these observables we however expects effects comparable to Higgs contributions, but suppressed by the larger $\eta$-mass.
Indeed, we find that the indirect bounds on the valid parameter points are very weak.
Therefore, we will restrict ourself to discussing only direct bounds on $\eta$ in the following, which we expect to have a significantly higher potential of probing the viable parameter points.

\begin{figure}
 \centering
 \includegraphics[keepaspectratio=true,width=0.48\textwidth]{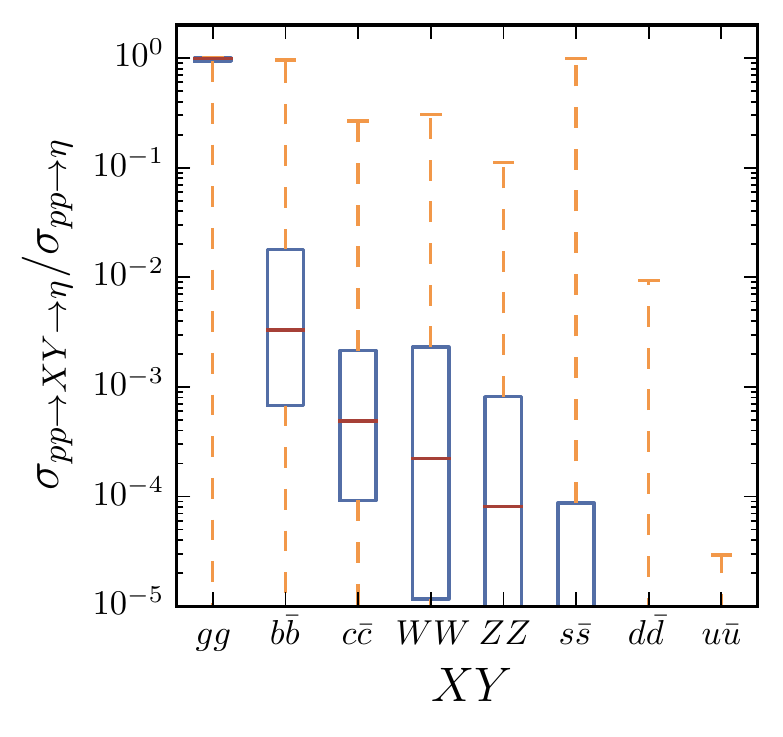}
 \quad
 \includegraphics[keepaspectratio=true,width=0.48\textwidth]{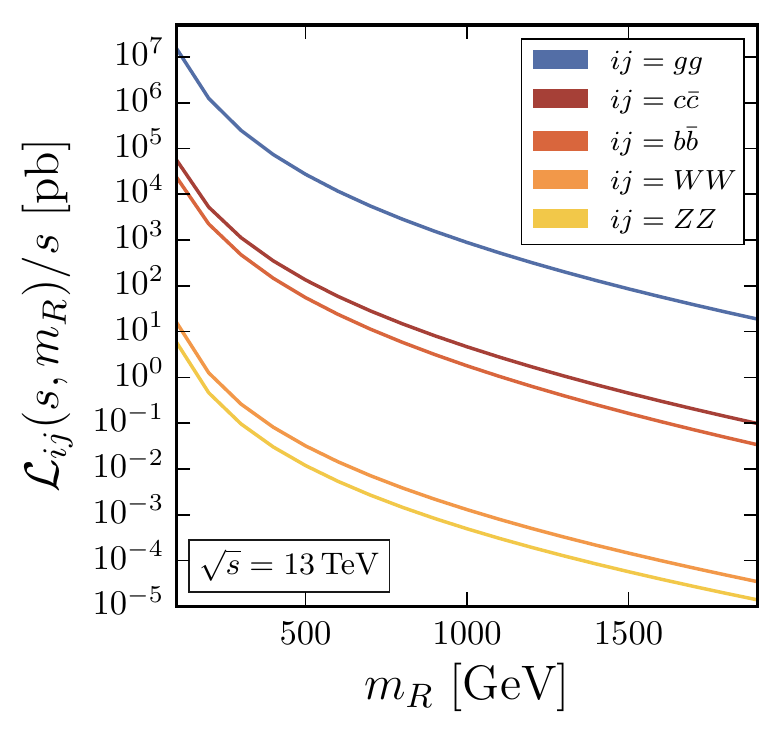}
 \caption{Left: Box plot of the $\eta$ production cross section in different channels relative to the total production cross section. For each channel we show the total range (indicated by dashed orange lines), the interquartile range (shown as a blue box) and the median (the red line inside the box) of values from viable parameter points.
 Right: Parton luminosities for $ij=gg,c\bar{c},b\bar{b}$ and effective parton luminosities from EWA for $ij=WW,ZZ$. We use $\sqrt{s}=13\,\text{TeV}$.}
 \label{fig:eta_production}
\end{figure}
Being a scalar that is allowed to mix with the Higgs boson, the production channels of $\eta$ are the same as for the Higgs.
It is thus expected that at a hadron collider like the LHC, the main channel is production via gluon fusion.
That this channel is indeed clearly dominating the total production cross section for most viable parameter points can be seen in figure~\ref{fig:eta_production}, where on the left the different production channels are compared in a box plot.
To be specific, for 50\% of viable points we find $r_{\sigma}(gg)=\sigma(pp\to gg\to \eta)/\sigma(pp\to\eta)>0.99$ and for half of the remaining points we still find $r_{\sigma}(gg)>0.93$. The main reason for this can be traced back to the parton luminosities entering the hadronic cross section in equation (\ref{eq:xsec}).
As shown in the right plot of figure~\ref{fig:eta_production}, the parton luminosity of gluons is ca. $10^2-10^3$ times larger than the one for bottom or charm quarks and even ca. $10^5-10^6$ times larger than the effective parton luminosities of W and Z bosons calculated through the EWA (cf. section \ref{sec:constraints}).
The light quarks on the other hand, while having a larger parton luminosity than bottom and charm quarks, still play a minor role in the $\eta$ production due to very small Yukawa couplings.

\begin{figure}
 \centering
 \includegraphics[keepaspectratio=true,width=0.459\textwidth]{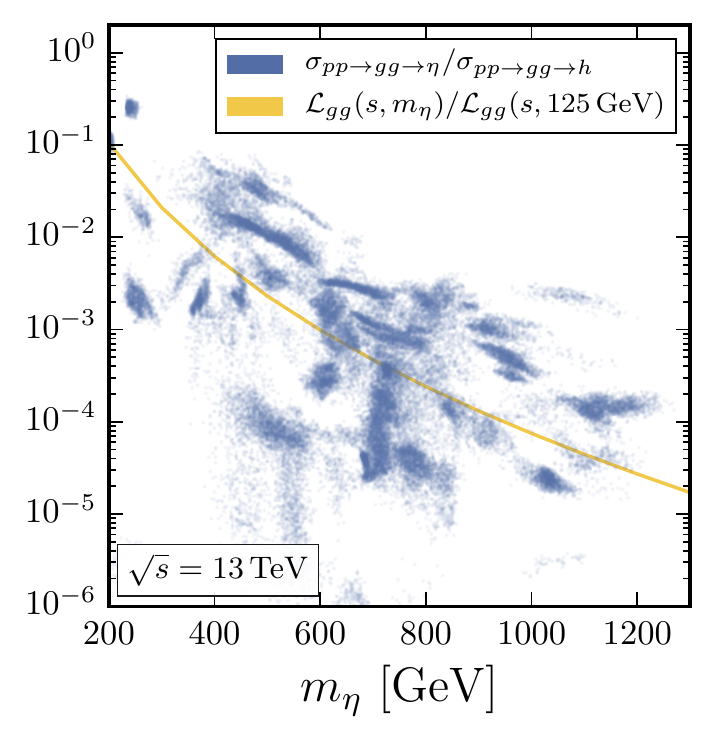}
 \quad
 \includegraphics[keepaspectratio=true,width=0.506\textwidth]{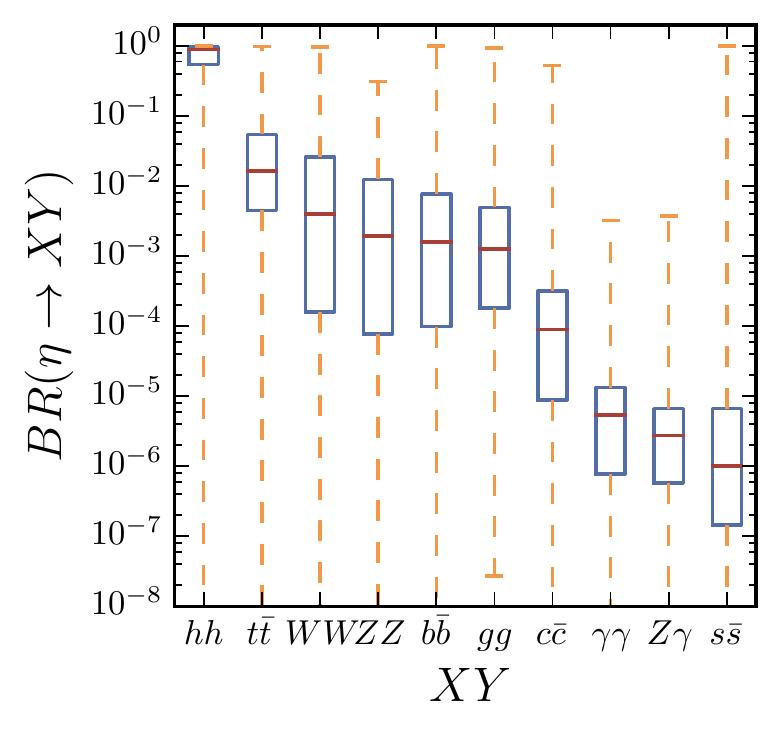}
 \caption{Left:
 Ratio of hadronic production cross sections via gluon fusion of $\eta$ and Higgs (blue dots) and ratio of gluon-gluon parton luminosities with $\sqrt{\hat{s}}=m_\eta$ and $\sqrt{\hat{s}}=125$~GeV.
 Right:
 Box plot of the $\eta$ branching ratio for different decay channels. We did not include the usually very small up and down quark branching ratios. Like in the left plot of figure~\ref{fig:eta_production}, we show the total and interquartile ranges and the median of values from viable parameter points.}
 \label{fig:eta_prod_decay}
\end{figure}

The gluon fusion cross section, which we have seen may serve as a good approximation of the total cross section, can on the partonic level be as large as the Higgs cross section, or even larger.
However, to get the hadronic cross section for gluon fusion, in the NWA we have to multiply the partonic one with the parton luminosity $\mathcal{L}_{gg}(s,\sqrt{\hat{s}})$.
This parton luminosity decreases by several orders of magnitude as the partonic center of mass energy $\sqrt{\hat{s}}$ grows from $125$~GeV to some hundred~GeV (cf. right plot in figure~\ref{fig:eta_production}).
The hadronic cross section of $\eta$ is thus suppressed compared to the one for the Higgs just because it has a higher mass and thereby a smaller parton luminosity entering equation (\ref{eq:xsec}).
A direct comparison of the values of $\mathcal{L}_{gg}(s,\sqrt{\hat{s}})$ at $\sqrt{\hat{s}}=m_\eta$ and $\sqrt{\hat{s}}=m_h=125$~GeV is shown in the left plot of figure~\ref{fig:eta_prod_decay}.
In the same plot, we show values of the ratio of the gluon fusion hadronic production cross sections of $\eta$ and Higgs for viable parameter points.
As expected, one observes that with increasing $\eta$ mass, the relative cross section decreases very similar to the relative parton luminosity.
There is still a broad range of possible values for the $\eta$ cross section at a given mass $m_\eta$, because the Yukawa couplings and fermion masses that enter the gluon fusion cross section may vary for different parameter points.
Nevertheless, also the maximum of possible values of the $\eta$ cross section decreases with larger $m_\eta$ and thus for the bulk of viable parameter points we get a suppression with respect to the Higgs cross section of at least $10^{-1}-10^{-2}$.

Like the production channels, the decay channels of $\eta$ are again the same as those of the Higgs -- with two important exceptions: if $m_\eta\ge2\,m_h$ and $m_\eta\ge2\,m_t$, which is the case for most viable parameters points, $\eta$ is allowed to decay to two Higgses or to $t\bar{t}$.
The $\eta hh$ coupling is calculated from the third derivatives of the two-dimensional scalar potential and leads to a large branching ratio $\text{BR}(\eta\to hh)$\footnote{%
The large $\eta hh$ coupling is a result of the $\gb{h}-\gb{\eta}$ mixing (cf. section \ref{sec:ScalarPotential}).
Since the electroweak Goldstone bosons yielding the longitudinal polarizations of $W$ and $Z$ do not mix with $\gb{\eta}$, the $\eta hh$ coupling is enhanced compared to the $\eta WW$ and $\eta ZZ$ couplings, contrary to what might be expected from the Goldstone boson equivalence theorem.}.
How large it actually is and that this leads to relatively small branching ratios in the remaining decay channels is shown in the box plot on the right side of figure~\ref{fig:eta_prod_decay}\footnote{We refrain from discussing decays to leptons. Due to the naive treatment of the lepton sector as beeing completely elementary, they have a strong dependence on the scalar mixing angle which we would not expect for a more complete model (cf. discussion in sec. \ref{sec:FermionSector}). We however note, that the branching ratios to leptons that we find are small enough to be neglected.}.
The usually second largest branching ratio is $\text{BR}(\eta\to t\bar{t})$ due to a mostly large Yukawa coupling.
While for the Higgs the decay to $b\bar{b}$ has the largest branching ratio, the $\eta$ can decay to two on-shell SM vector bosons for most of the viable parameter points and thus $\text{BR}(\eta\to WW)$ and $\text{BR}(\eta\to ZZ)$ is usually larger than $\text{BR}(\eta\to b\bar{b})$.
The smallest branching ratios are found for the loop-induced decays involving massless vector bosons and for those to light quarks with small Yukawa couplings.

\begin{figure}
 \centering
 \includegraphics[keepaspectratio=true,width=0.48\textwidth]{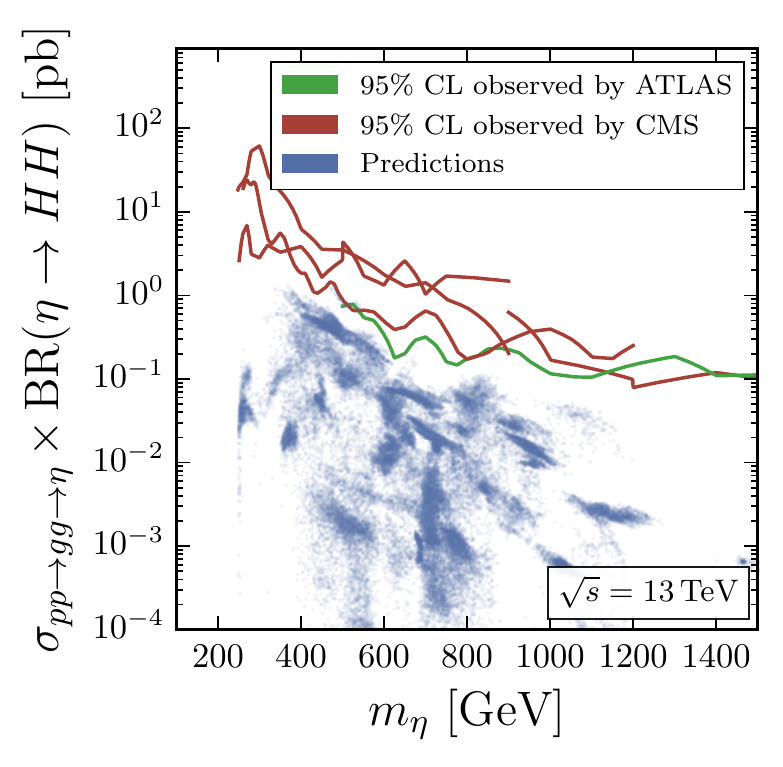}
 \quad
 \includegraphics[keepaspectratio=true,width=0.48\textwidth]{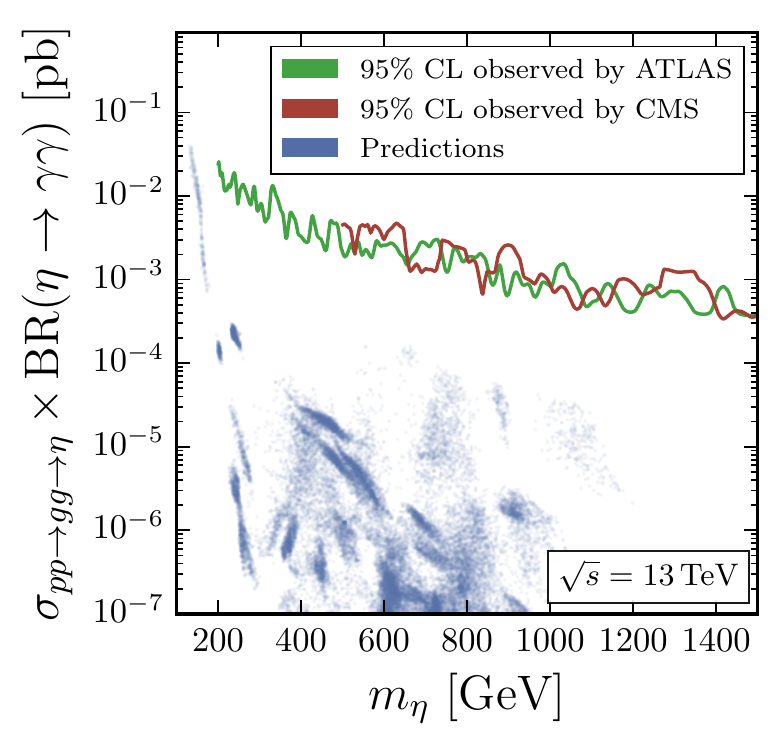}
 \\
 \includegraphics[keepaspectratio=true,width=0.48\textwidth]{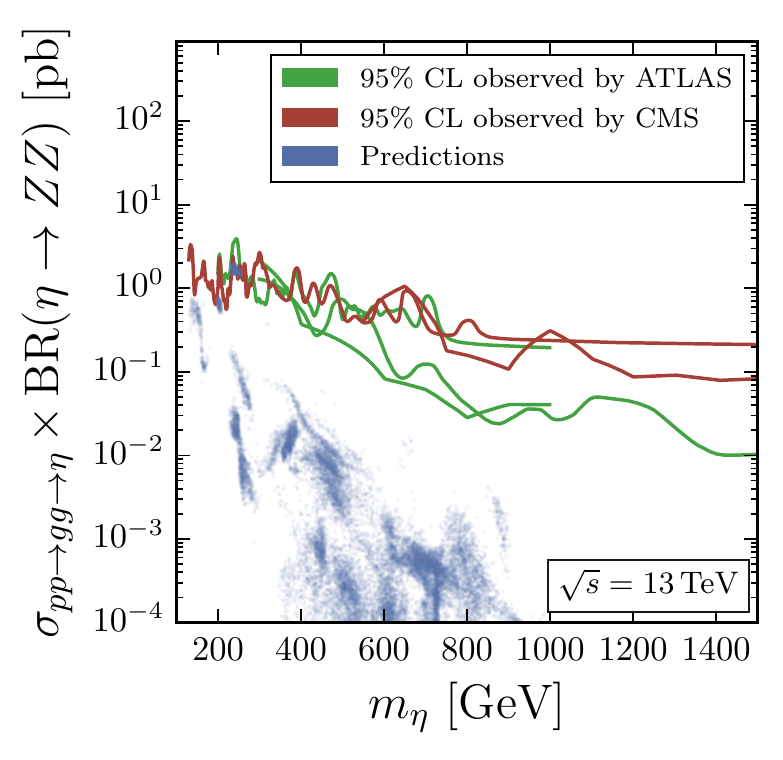}
 \quad
 \includegraphics[keepaspectratio=true,width=0.48\textwidth]{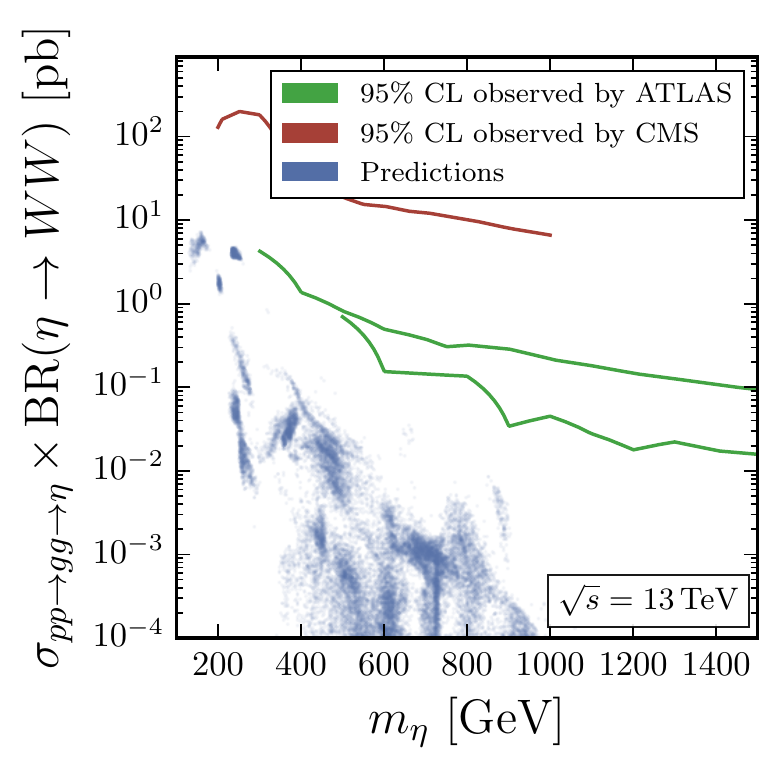}
 \caption{Experimental bounds from ATLAS and CMS and predictions from viable parameters points of the $\eta$ production cross section via gluon fusion times the branching ratio into $hh$ (top-left), $\gamma\gamma$ (top-right), $ZZ$ (bottom-left) and $WW$ (bottom-right). The analyses shown in the plots are listed in table \ref{tab:exp_scalar_res}.}
 \label{fig:eta_bounds}
\end{figure}
Since we have implemented many experimental analyses into our scanning procedure (cf. tables \ref{tab:exp_quark_res}, \ref{tab:exp_scalar_res} and \ref{tab:exp_vector_res}), we are able to compare the experimental bounds on the product of cross section and branching ratio into different channels to the predictions from the viable parameters points.
As expected from the discussion of $\eta$ branching ratios above, the decay channel to two Higgses should be the most promising one for setting bounds.
This can indeed be observed in the upper-left plot of figure~\ref{fig:eta_bounds}.
Even though the decay to two Higgses is not the easiest one to observe experimentally, the ATLAS and CMS collaborations are able to exclude a cross section times branching ratio at 95\% CL in this channel down to $10^0-10^{-1}$~pb.
This is very close to the values we predict for many of the viable parameter points.
We thus expect that this channel has a high potential to exclude parts of these points already during the current LHC run 2.
Another channel with good prospects is the decay to two $Z$ bosons, for which we show a plot on the lower-left of figure~\ref{fig:eta_bounds}.
While the predicted values for cross section times branching ratio are smaller than for the decay $\eta\to hh$, the experimental collaborations are able to also exclude smaller values in this channel.
Actually, the experimental bounds are strong enough that some still viable parameter points with low $\eta$ mass are literally on the verge of being excluded.
Due to the custodial symmetry, the predictions in the $\eta\to ZZ$ channel are very similar to the decay $\eta\to WW$, for which we show limits and predictions in the lower-left plot of figure~\ref{fig:eta_bounds}.
While at the moment there are less experimental analyses available for this channel than for $\eta\to ZZ$, both are promising.
On the upper-right we show bounds in the diphoton channel from LHC Run 2 data.
Even though the experiments are able to probe values down to less than $10^{-3}$~pb in this channel, the bulk of our predicted values are still far from beeing excluded.
Apart from very few cases, the points with the highest signal strength are around two orders of magnitude away from the experimental bounds.
The main reason for this is the tiny branching ratios we find for $\eta\to\gamma\gamma$ (cf. right plot in figure~\ref{fig:eta_prod_decay}).
The $\eta\to Z\gamma$ channel, for which no plot is shown, is very similar to the $\eta\to\gamma\gamma$ case.
While the decays to light quarks can be neglected due to the tiny branching ratios, the decays of $\eta$ to third generation quarks or gluons have a relatively large branching ratio. However, the experimental bounds in these channels are even more far away from excluding viable parameter points than those in the diphoton channel and thus we also refrain from showing plots for these channels.

We have seen that the LHC experiments may probe parts of the still viable parameter space in the near future.
However, we expect many analyses to aim at setting limits for as high as possible resonance masses while cutting out the lower mass range.
Since $\eta$ has a mass well below $1$~TeV for many of the viable parameter points, we want to stress the importance of also exploring this mass range with higher luminosity.

\subsection{Phenomenology of vector and fermion resonances}

\begin{figure}
 \centering
 \includegraphics[keepaspectratio=true,width=0.48\textwidth]{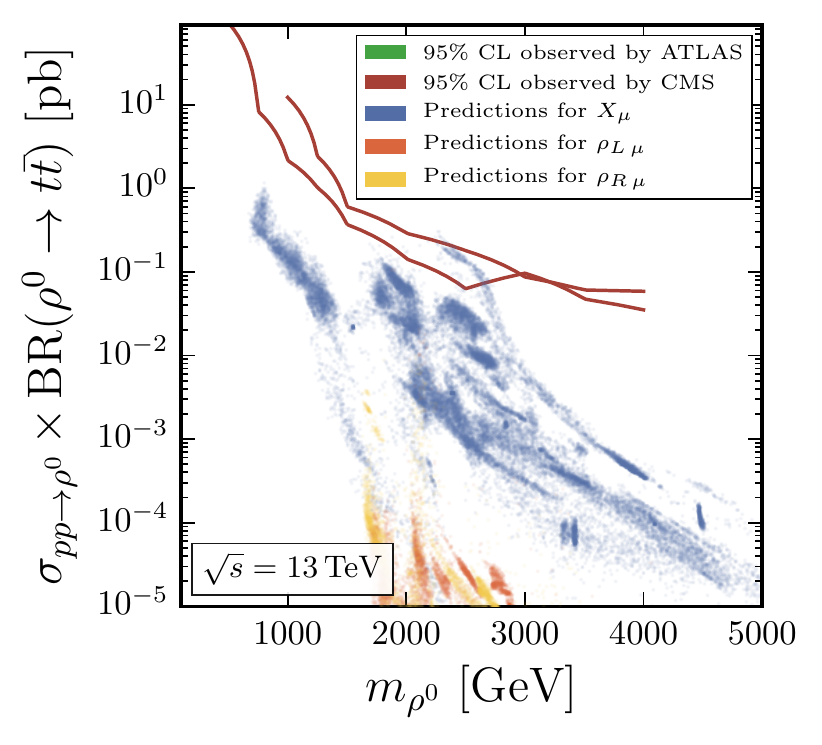}
 \quad
 \includegraphics[keepaspectratio=true,width=0.48\textwidth]{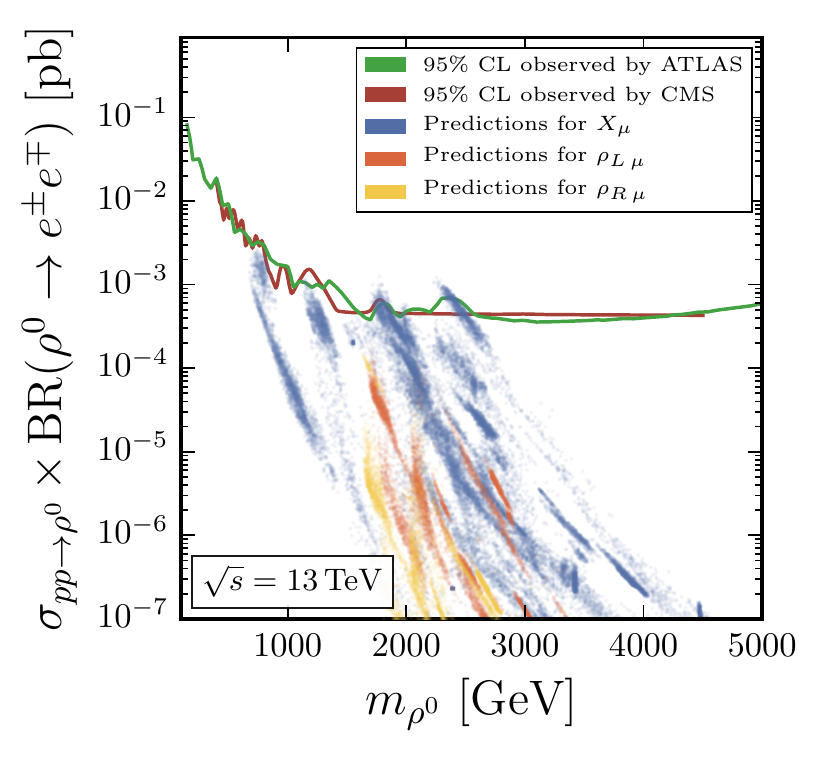}
 \\
 \includegraphics[keepaspectratio=true,width=0.48\textwidth]{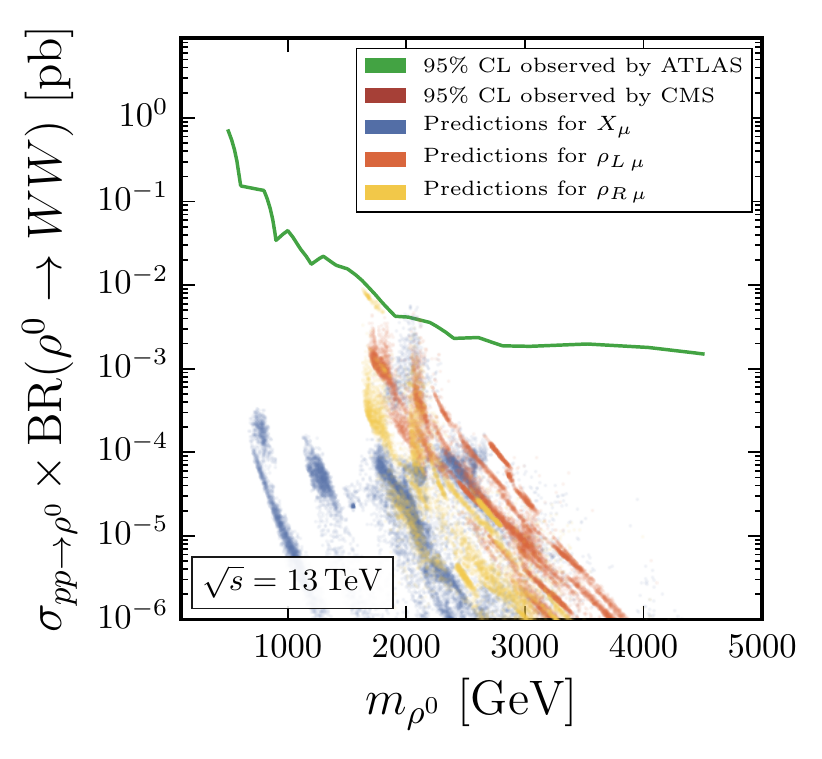}
 \quad
 \includegraphics[keepaspectratio=true,width=0.48\textwidth]{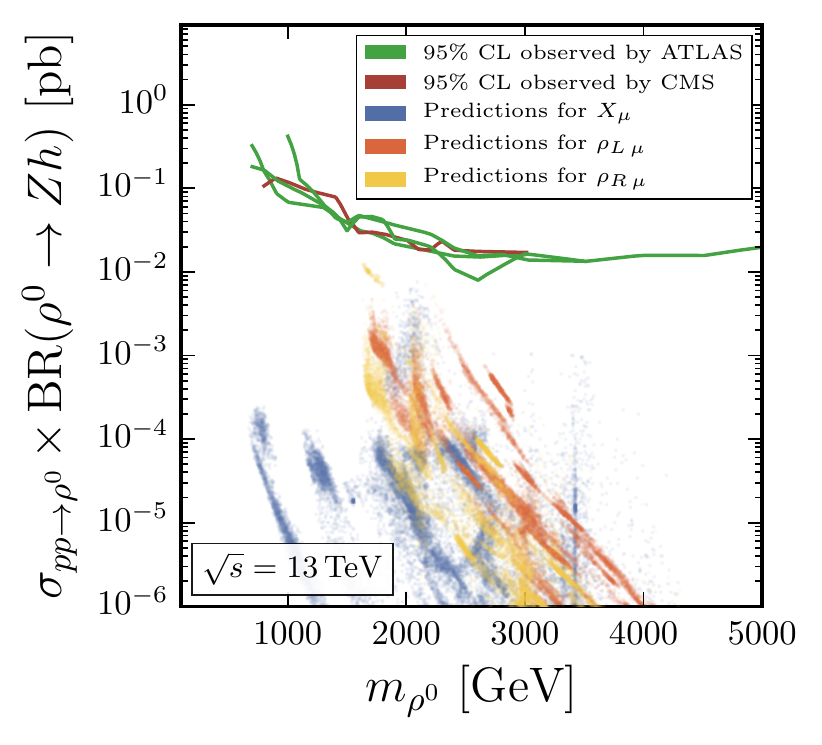}
 \caption{Experimental bounds from ATLAS and CMS and predictions for the neutral vector resonance production cross section times the branching ratio into $t\bar{t}$ (top-left), $e^+ e^-$ (top-right), $WW$ (bottom-left) and $ZH$ (bottom-right).
 We show values for the resonances $X_\mu$, $\rho_{L\,\mu}$ and $\rho_{R\,\mu}$ for the viable parameter points.
 The analyses included in the plots are listed in table \ref{tab:exp_vector_res}.}
 \label{fig:rho0_bounds}
\end{figure}

While the main features and properties of the vector and fermion resonances are still the same as in the model analysed in~\cite{Niehoff:2015iaa}, some additional states are introduced due to the now larger spontaneously broken global symmetry SO$(6)$ instead of SO$(5)$\footnote{An $\text{SO}(6)/\text{SO}(5)$ composite Higgs model with essentially the same electroweak spin-1 resonance states as presented here was discussed in \cite{Franzosi:2016aoo}. However, because this model does not include partial compositeness, the couplings of SM quarks to spin-1 resonances are different to those of the model considered here.}.
Furthermore, a lot of new experimental analyses based on LHC run 2 data with center-of-mass energy of $13$~TeV have been published by ATLAS and CMS during the last year that have not been available for the discussion in~\cite{Niehoff:2015iaa}.
In the following, we will thus present the most promising decay channels for constraining the heavy resonances after taking into account the new experimental data.

\subsubsection{Prospects for vector resonance searches}\label{sec:bounds_vectors}
Like discussed in section \ref{sec:BosonSector}, we consider charged and neutral electroweak resonances as well as a coloured gluon resonance $\mathcal{G}_\mu$.
The latter is always heavy enough to decay to a pair of quark resonances and thus gets very broad.
Therefore, to set bounds on the gluon resonance, it is most promising to search for the decay of the pair of quark resonances \cite{Azatov:2015xqa,Araque:2015cna}. It is however beyond the scope of this analysis to take these effects into account.
Since additionally the branching ratios to SM quarks are very small, there are effectively no bounds on the gluon resonance in our numerical analysis.
In the following, we thus focus on the electroweak vector boson sector which contains eight neutral and four charged resonances.

\paragraph{Neutral electroweak resonances}
As expected from the vector boson mass matrix (cf. appendix \ref{appendix:MassMatrices:Vector}), out of the eight heavy neutral mass eigenstates\footnote{The mixing in the spin-1 sector is moderate such that each mass eigenstate can be associated to an eigenstate in the gauge basis. We thus denote each mass eigenstate by the name of the gauge basis state it is mainly composed of.}, we observe four states to be always relatively light (corresponding to $\rho_{L\,\mu}$, $\rho_{R\,\mu}$, $a_{1\,\mu}^3$ and $a_{1\,\mu}^4$) and three states to be always heavy ($\rho_{S\,\mu}$, $a_{2\,\mu}^3$ and $a_{2\,\mu}^4$).
The resonance $X_{\mu}$ has a mass independent of the other resonances. It can thus be the lightest one, can have a mass between the four light and three heavy states or can also be the heaviest one.
In the case where $X_{\mu}$ is the lightest resonance, it may even be lighter than the naive lower bound on vector resonance masses from the $S$~parameter might suggest.
This is due to the fact that a KK photon like linear combination of neutral electroweak resonances does not contribute to the $S$~parameter. For $g_X \ll g_\rho$ (and thus a small $X_\mu$ mass), the KK photon like linear combination is mainly composed of $X_{\mu}$, which therefore can be very light and still compatible with the bound from the $S$~parameter (cf. discussion in~\cite{Niehoff:2015iaa}).

Compared to the $\text{SO}(5)/\text{SO}(4)$ model there are three additional neutral resonances: the SU$(2)_L\times$SU$(2)_R$ singlet $\rho_{S\,\mu}$ and the two neutral components $a_{1\,\mu}^3$ and $a_{1\,\mu}^4$ of the SU$(2)_L\times$SU$(2)_R$ bidoublet $a_{1\,\mu}$. While $\rho_{S\,\mu}$ and $a_{1\,\mu}^4$ do not mix with $W_\mu^3$ and $B_\mu$, there is a mixing term proportional to $\tilde{s}^*_\eta$ for $a_{1\,\mu}^3$ (cf. appendix \ref{appendix:MassMatrices:Vector}).
In the limit where one recovers the $\text{SO}(5)/\text{SO}(4)$ case and $\tilde{s}^*_\eta\to0$ (cf. discussion in section \ref{sec:ScalarPotential}), the mixing term of $a_{1\,\mu}^3$ thus vanishes.
At the same time, the mixing terms of all other resonances take the form of the $\text{SO}(5)/\text{SO}(4)$ case with $a_{1\,\mu}^3$ and $a_{1\,\mu}^4$ playing the role of the states that were denoted by $\mathfrak{a}_{\mu}^3$ and $\mathfrak{a}_{\mu}^4$ in~\cite{Niehoff:2015iaa}.
Due to the large mass and the absence of mixing for $\rho_{S\,\mu}$ and $a_{1\,\mu}^4$, the only new state that is potentially relevant for the phenomenology is $a_{1\,\mu}^3$.
However, we observe that $a_{1\,\mu}^3$ always has a very small production cross section.
Consequently, it can be neglected when discussing the collider phenomenology and the differences to the $\text{SO}(5)/\text{SO}(4)$ case mainly arise due to the dependence of the mixing terms on $\tilde{s}^*_\eta$.

The relevant states that have a significant production cross section are $\rho_{L\,\mu}$, $\rho_{R\,\mu}$ and $X_{\mu}$.
For the case that $X_{\mu}$ is the lightest resonance, it has the by far highest production cross section of all resonances.
Therefore, this case is the most strongly constrained one.
Since $X_{\mu}$ can couple to $h$ only via mixing, its coupling to $Zh$ is very small.
Additionally, the $X_{\mu}$ gauge eigenstate only mixes with $B_\mu$ but not with $W_\mu^3$ and thus the coupling of $X_{\mu}$ to $WW$ is also strongly suppressed.
$X_{\mu}$ can couple to leptons, but again only via mixing with the $B_\mu$.
The largest branching ratio of a light $X_{\mu}$ to SM particles\footnote{If
kinematically allowed, $X_{\mu}$ can also decay to fermion resonances.
In this case, the branching ratios to SM particles decrease and at the same time the resonance becomes very broad such that it might not be captured by experimental analyses.
To take this into account, we relax the experimental bounds for broad resonances by multiplying the corresponding $\chi^2$ value with a smooth function that is close to one for small $\Gamma/m$ and vanishes for large $\Gamma/m$.} is thus found for quarks in the final state, in particular for the $X_{\mu}\to t\bar{t}$ channel.
In the upper-left plot of figure~\ref{fig:rho0_bounds} we show that this leads to large values of cross section times the branching ratio for many parameter points.
The experimental bounds are already in the region of predicted values\footnote{Viable parameter points in our scan are allowed to violate experimental bounds by up to three sigma. Since the experimental limits are on the 95\% CL, we find points that lie above the experimental bounds that are shown in the plots.} and we expect still viable parameter points to be probed in this channel already during the current LHC run 2.
Even though we do not consider partial compositeness in the lepton sector and thus the spin-1 states in the composite sector can couple to leptons only via their mixing with the elementary ones, the bounds from resonances decaying to two leptons are even stronger than in the $t\bar{t}$ channel.
The reason for this is the ability of the experiments to probe much lower values of cross section times branching ratio in the di-electron and di-muon channels than they can probe for the $t\bar{t}$ final state.
In the upper-right plot of figure~\ref{fig:rho0_bounds} one can observe that many of the still viable parameter points are closely adjacent to the exclusion limits in the di-electron channel and thus expected to be probed in the near future.
This is mainly due to the $X_{\mu}$ that can have a much larger production cross section than the $\rho_{L\,\mu}$ and the $\rho_{R\,\mu}$ if it is the lightest resonance.
In the case where the $X_{\mu}$ is heavier than $\rho_{L\,\mu}$ and $\rho_{R\,\mu}$, the latter can play the most important role in probing the parameter space.
In contrast to $X_{\mu}$, the resonances $\rho_{L\,\mu}$ and $\rho_{R\,\mu}$ have couplings to SM dibosons that are not strongly suppressed.
So apart from the then still promising dilepton channels, the $WW$ and $Zh$ channels are also very interesting in this case.
The plot in the lower-left of figure~\ref{fig:rho0_bounds} shows that especially in the $WW$ channel some still viable parameter points are not far away from current experimental bounds.
While we predict similarly high values of cross section times branching ratio in the $Zh$ channel (cf. lower-left plot of figure~\ref{fig:rho0_bounds}) and more experimental analyses are available there, the bounds are still farther away than in the $WW$ channel.

In summary, while for the case of $X_{\mu}$ being the lightest resonance, the $t\bar{t}$ and dilepton channels are the most promising ones, in the case of $\rho_{L\,\mu}$ and $\rho_{R\,\mu}$ being lighter than $X_{\mu}$, the dilepton and the $WW$ channels have the highest prospects of observing or excluding parameter points.

\paragraph{Charged resonances}
\begin{figure}
 \centering
 \includegraphics[keepaspectratio=true,width=0.48\textwidth]{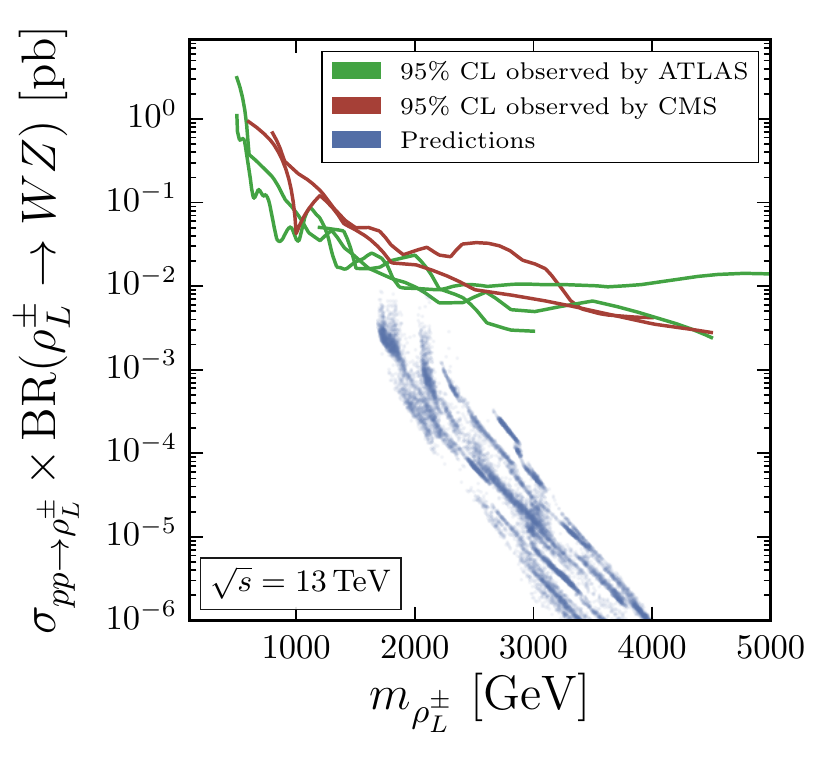}
 \quad
 \includegraphics[keepaspectratio=true,width=0.48\textwidth]{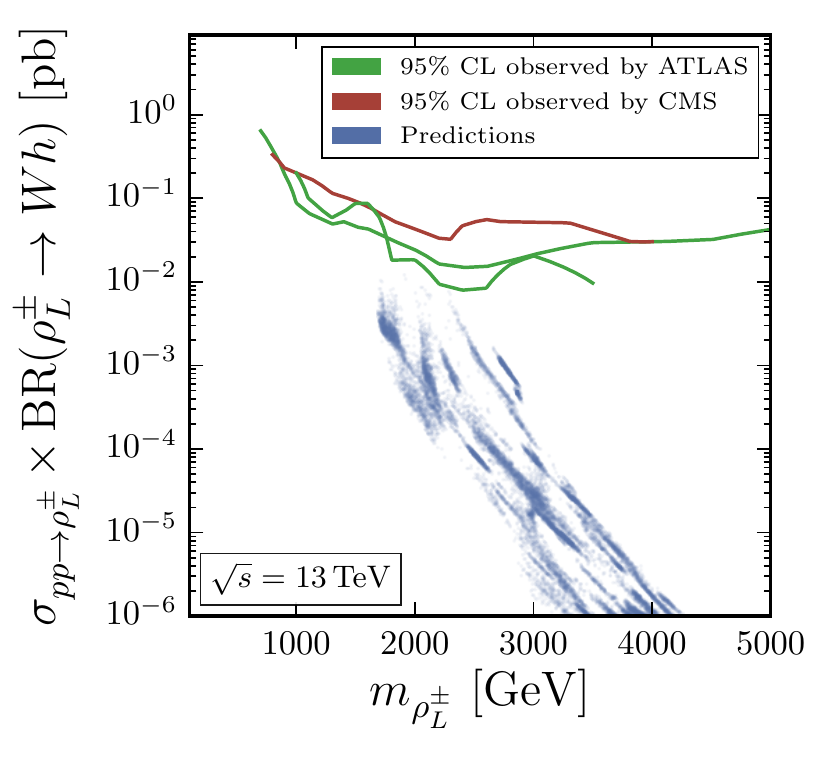}
 \\
 \includegraphics[keepaspectratio=true,width=0.48\textwidth]{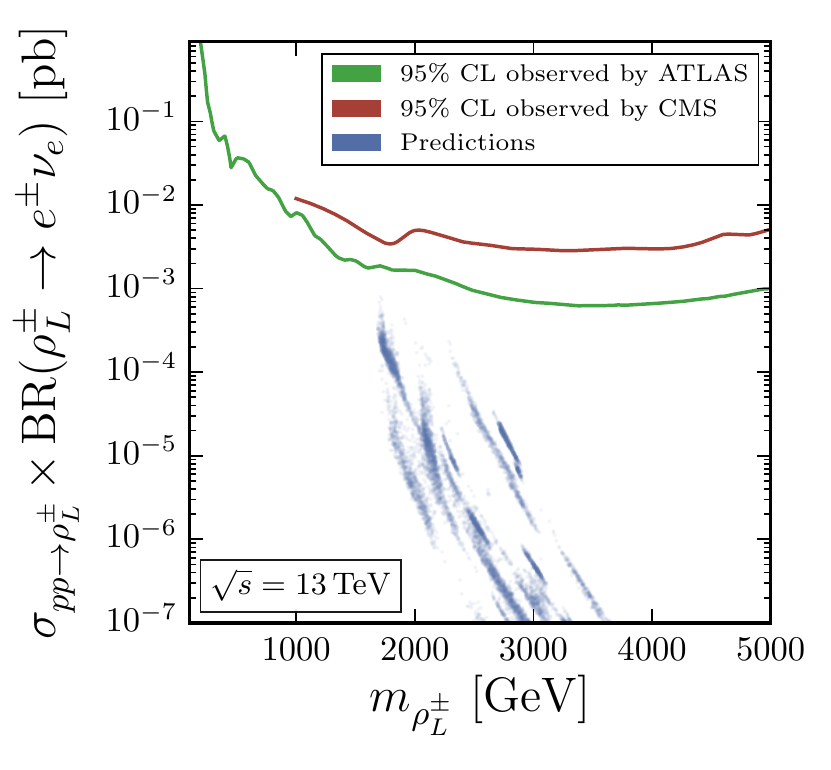}
 \quad
 \includegraphics[keepaspectratio=true,width=0.48\textwidth]{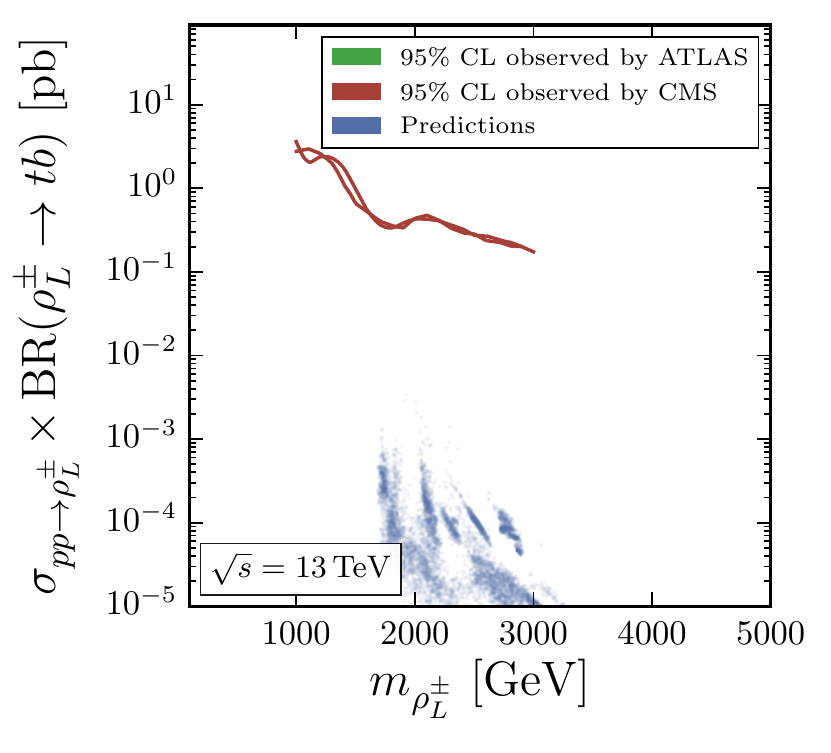}
 \caption{Experimental bounds from ATLAS and CMS and predictions from viable parameters points of the $\rho_{L\,\mu}^\pm$ production cross section times the branching ratio into $WZ$ (upper-left), $Wh$ (upper-right), $e^\pm\nu_e$ (lower-left) and $tb$ (lower-right). The analyses shown in the plots are listed in table \ref{tab:exp_vector_res}.}
 \label{fig:rhopm_bounds}
\end{figure}

Among the four charged vector resonances, three of them have very similar masses ($\rho_{L\,\mu}^\pm$, $\rho_{R\,\mu}^\pm$ and $a_{1\,\mu}^\pm$) and one is always heavier ($a_{2\,\mu}^\pm$).

The state $a_{1\,\mu}^\pm$ is the only one not present in the $\text{SO}(5)/\text{SO}(4)$ model discussed in~\cite{Niehoff:2015iaa}. Similarly to the discussion for the neutral resonances, taking $\tilde{s}^*_\eta\to0$ yields a vanishing mixing term for $a_{1\,\mu}^\pm$, all other mixings take the form of the $\text{SO}(5)/\text{SO}(4)$ case and $a_{2\,\mu}^\pm$ plays the role of the state that was called $\mathfrak{a}_{\mu}^\pm$ in~\cite{Niehoff:2015iaa}.

When it comes to the collider phenomenology, it suffices to discuss the effects of $\rho_{L\,\mu}^\pm$ since it always has a considerably higher production cross section than $\rho_{R\,\mu}^\pm$ and $a_{1\,\mu}^\pm$.
The highest branching ratios of $\rho_{L\,\mu}^\pm$ to SM particles are found in the $WZ$ and $Wh$ channels.
We show predicted values for production cross section times branching ratio in these channels in the two upper plots of figure~\ref{fig:rhopm_bounds}.
While the branching ratio to $Wh$ is usually slightly larger than the one to $WZ$, the experiments have a bit more sensitivity in the latter case.
For both channels we find points that are in reach of near future experimental bounds.
While the predictions in the $e^\pm\nu_e$ channel (cf. lower-left plot in figure~\ref{fig:rhopm_bounds}) are smaller than those in the diboson channels by at least a factor of $10$, due to the higher sensitivity of experimental analyses they are expected to also probe still viable parameter points in this channel during LHC run 2.
We refrain from showing a plot for the decay to $\mu^\pm\nu_\mu$ since it is very similar to the one for the $e^\pm\nu_e$ channel.
In the $\rho_{L\,\mu}^\pm\to tb$ channel on the other hand (cf. lower right plot in figure~\ref{fig:rhopm_bounds}), the predictions are far away from the experimental bounds such that even with a lot of additional integrated luminosity this channel is not very promising.

\subsubsection{Prospects for quark resonance searches}
\begin{figure}
 \centering
 \includegraphics[keepaspectratio=true,width=0.48\textwidth]{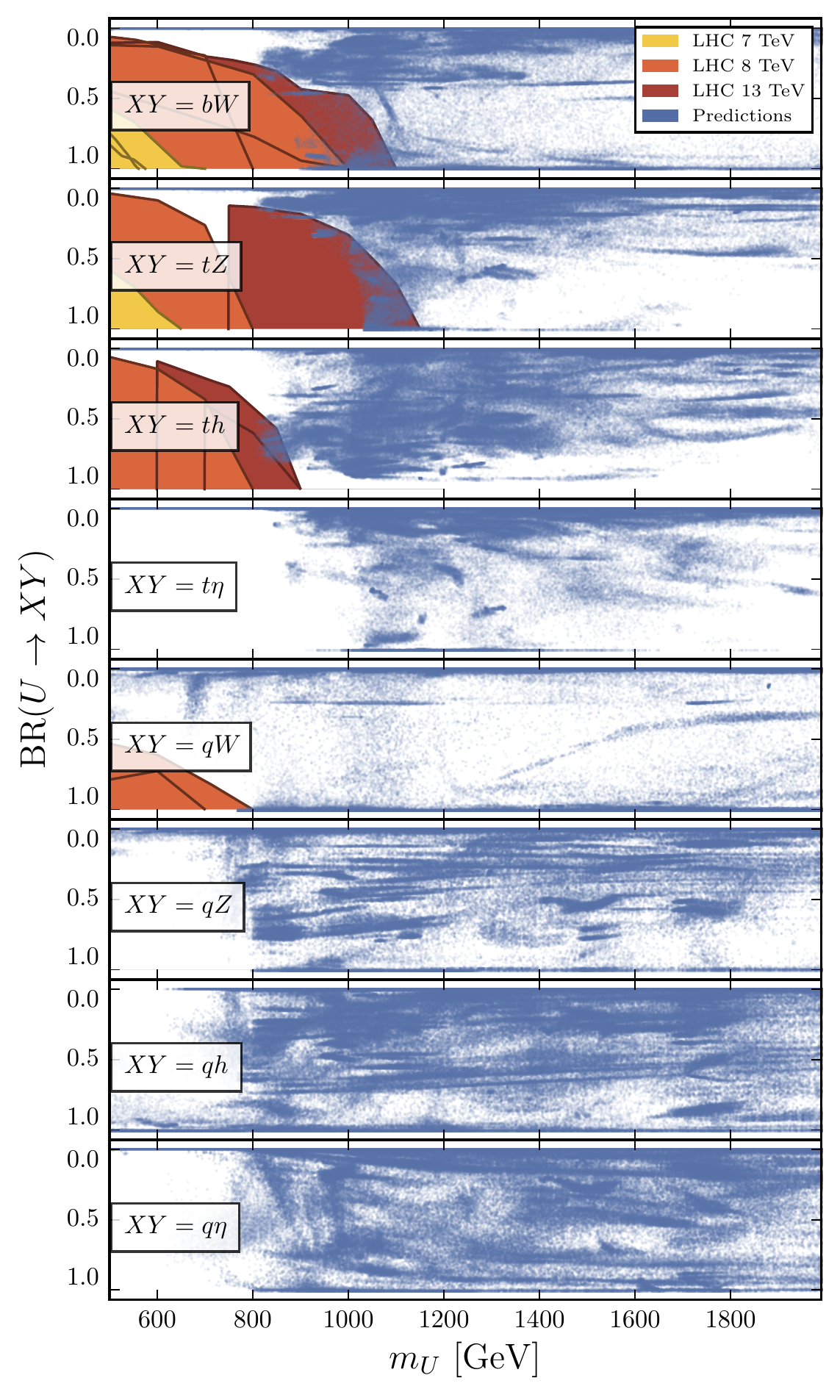}
 \quad
 \includegraphics[keepaspectratio=true,width=0.48\textwidth]{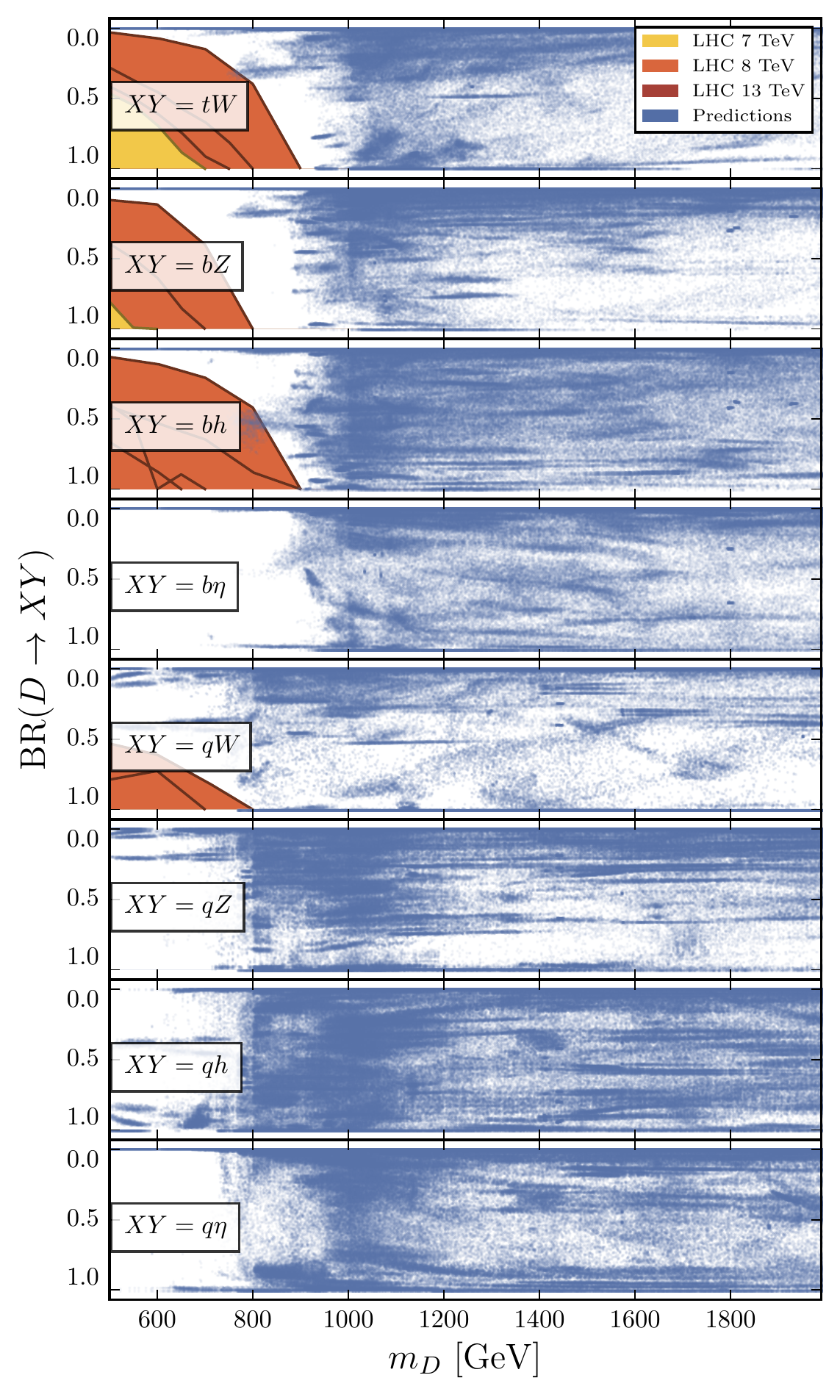}
 \\
 \includegraphics[keepaspectratio=true,width=0.48\textwidth]{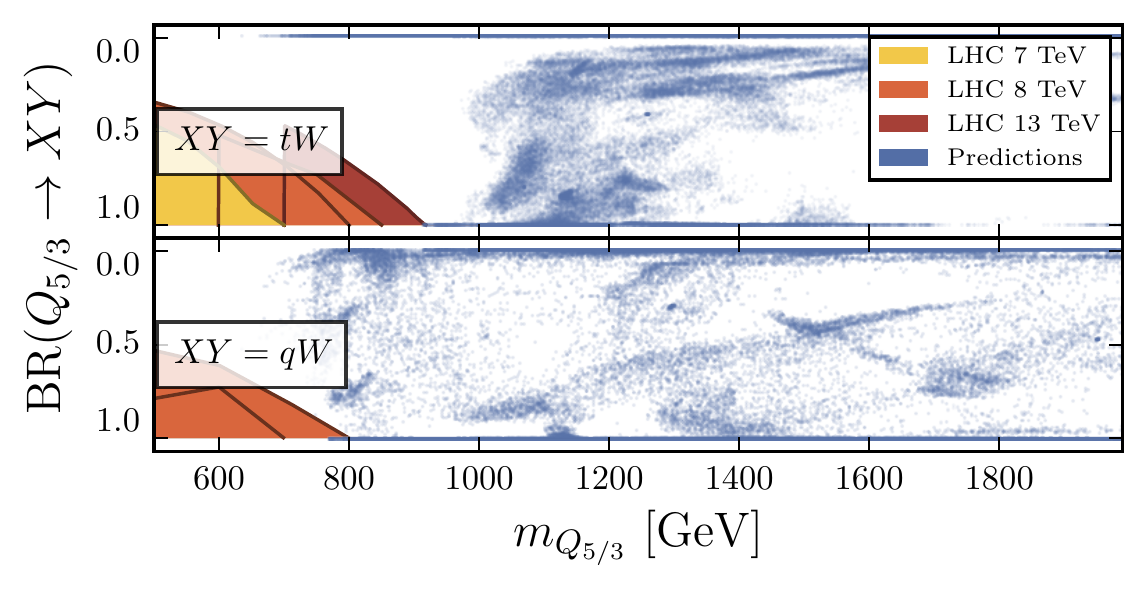}
 \quad
 \includegraphics[keepaspectratio=true,width=0.48\textwidth]{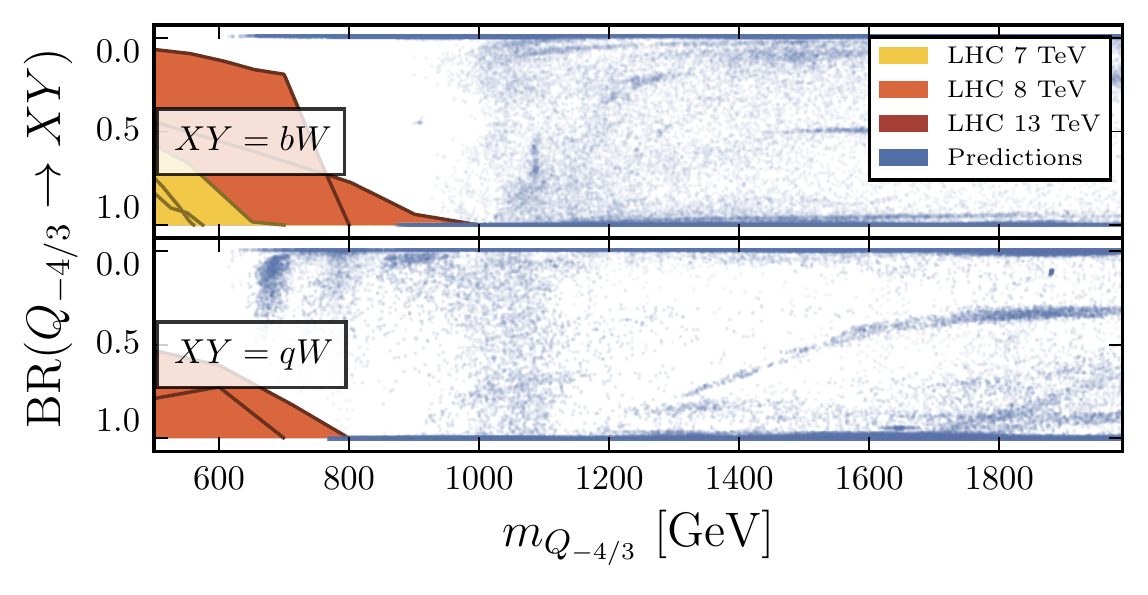}
 \caption{Predictions for masses and branching ratios of up-type (upper-left), down-type (upper-right), charge $5/3$ (lower-left) and charge $-4/3$ (lower-right) quark resonances.
 Experimental bounds from the LHC running at different center of mass energies are shown as coloured areas. The analyses that are included in the plots are listed in table \ref{tab:exp_quark_res}.}
 \label{fig:Q_bounds}
\end{figure}
In contrast to the spin-1 resonance case, there is usually strong mixing in the quark sector.
Therefore, we refrain from identifying the gauge eigenstate from which each mass eigenstate is mainly composed of and simply discuss all quark resonances at once.
Among the 30 states, the most interesting ones for the collider phenomenolgy are the lightest ones that always decay to a final state containing a SM quark and can thus be directly searched for at the LHC.
For setting bounds, we consider the model-independent pair production of quark resonances\footnote{As already mentioned in section \ref{sec:bounds_vectors}, we neglect possible contributions to the quark resonance pair production stemming from the gluon resonance.

While many recent experimental analyses focus on single production to reach a higher sensitivity, it is not feasible to use their model dependent bounds in our numerical analysis.}.
Experimental limits on cross section times branching ratio can thus be easily recast into bounds on the branching ratio of a given decay channel, which only depend on the mass of the resonance.
Observed 95\% CL bounds in several channels are shown as coloured areas in figure~\ref{fig:Q_bounds}.
For all viable parameter points, we show predicted values for masses and branching ratios of all quark resonances lighter than 2 TeV.
A common feature of all plots in figure~\ref{fig:Q_bounds} is that the branching ratios usually decrease with higher masses since new decay channels to other quark resonances open up.
The most interesting region is thus the one around 1 TeV, where most resonances have high branching ratios to SM particles.
We want to note, that we find at least one quark resonance with a mass below 1.2 TeV for 97\% of the viable parameter points.
Probing this mass region by direct searches for quark resonances has thus arguably the highest potential to observe or exclude our viable parameter points with LHC run~2 data.

The currently strongest experimental bounds are on the decays of up-type quark resonances~$U$ with a third generation SM quark in the final state (cf. upper-left plot in figure~\ref{fig:Q_bounds}).
In contrast to~\cite{Niehoff:2015iaa}, where the analyses published during the last year have not yet been available, many parameter points are closely adjacent to or even above the observed 95\% CL\footnote{Like in section \ref{sec:bounds_vectors}, the experimental limits shown in the plots are on the 95\% CL, while we allow viable parameter points to violate experimental bounds by up to three sigma.}.
While also new analyses for decays of down-type resonances~$D$ have appeared since the publication of~\cite{Niehoff:2015iaa}, none of them already features data from LHC run~2 and thus the current bounds on $D$~resonances (cf. upper-right plot in figure~\ref{fig:Q_bounds}) are much weaker than those on~$U$.

Since we consider a model with partial compositeness for all three quark generations, especially the lightest resonances that we find are dominantly decaying to final states involving a light SM quark.
Apart from a few searches with a $W$~boson and a light quark in the final state, there are essentially no direct experimental bounds on quark resonances decaying to light quarks.
Interestingly, we find essentially all quark resonances with a mass below 750~GeV to be mainly composed of the singlets $S^2$ and $\tilde{S}^2$, which in this case dominantly decay to a light SM quark and a Higgs (cf. ``$XY$~$=$~$qh$''-channel in upper plots of figure \ref{fig:Q_bounds}).
This channel is thus by far the most promising one to set direct bounds on the still unconstrained very light quark resonances found in our scan.

In the lower plots of figure~\ref{fig:Q_bounds}, we show predictions and experimental bounds for quark resonances with exotic charges $5/3$ and $-4/3$.
For nearly all viable parameter points, the exotic charged resonances are heavier than the lightest up- or down-type resonance.
While the mixing with SM quarks can lower the mass for up- and down-type resonances, this is not possible for the exotic charged ones.
Apart from decaying to SM particles, the lightest exotic resonances are thus also allowed to decay to other resonances, which slightly weakens the bounds one can set.
Nevertheless, we find many viable parameter points predicting masses and branching ratios of exotic charged quarks that are in reach of LHC run 2.

In addition to the decays featuring a SM boson in the final state, also decays to a SM quark and the scalar singlet $\eta$ are allowed\footnote{A discussion of quark resonance phenomenology featuring decays to the singlet $\eta$ is presented in~\cite{Serra:2015xfa}. The model discussed there does however not contain mixing in the scalar sector and only third-generation partners are considered.}.
We show predictions for up- and down-type resonances decaying to $\eta$ and SM quarks in figure~\ref{fig:Q_bounds}.
While the presence of the additional decay channel may obviously alter the branching ratios in the other channels, due to our assumption of partial compositeness for all three generations, there are already many channels the quark resonances can decay to.
Consequently, the overall picture is not changed very much to what is observed in~\cite{Niehoff:2015iaa}, where this channel is not available.

\subsection{Flavour physics and $C\! P$ violation}

Constraints from flavour physics are a well-known challenge for composite Higgs
models, even with partial compositeness. Since we consider a model with a
flavour symmetry in the strong sector only broken by composite-elementary
mixings, these constraints can be under control, but are still relevant.
Since the effects in flavour physics are mostly dependent on the flavour
structure and not so much on the chosen coset,
the flavour effects are similar to the results found for the
$\text{SO}(5)/\text{SO}(4)$ coset with a $\text{U}(2)^3_\text{RC}$ flavour
symmetry presented in~\cite{Niehoff:2015iaa}.
However, the presence of spontaneous $C\! P$ violation in the scalar sector leads
to additional flavour blind $C\! P$ violation that can manifest itself in
electric dipole moments.

\subsubsection{Meson-antimeson mixing}

The predictions for mixing observables in the $K$, $B_d$, and $B_s$ systems
turn out to be qualitatively very similar to the
$\text{SO}(5)/\text{SO}(4)$ coset with a $\text{U}(2)^3_\text{RC}$ flavour
symmetry. We repeat the main predictions of this scenario.
\begin{itemize}
 \item Modifications of the mass differences $\Delta M_d$ and $\Delta M_s$
 can saturate current experimental bounds.
 Relative to the SM, for fixed values of CKM elements, the mass differences
 are always \textit{enhanced}.
 Future improvements will require
 reducing the parametric uncertainties of the SM predictions by more precise
 determinations of the CKM quantities $|V_{cb}|$, $|V_{ub}|$, and $\gamma$ from
 tree-level $B$ decays, as well as of the relevant matrix elements from lattice QCD.
 \item Indirect $C\! P$ violation in kaon mixing measured by the parameter $\epsilon_K$
 can also saturate present bounds.  Relative to the SM, for fixed values of
 CKM elements, $|\epsilon_K|$ is always \textit{enhanced} and the relative
 enhancement is \textit{always larger} than in the $B_{d,s}$ mass differences.
 \item The $B_s$ mixing phase $\phi_s$ can receive a new physics contribution
 up to $\pm0.1$, i.e.\ in the ball park of the present experimental bounds,
 arising from subleading terms in the $\text{U}(2)^3$ spurion expansion.
 $C\! P$ violation in $B_d$ mixing is instead SM-like.
\end{itemize}

\subsubsection{Rare $B$ decays}
\begin{figure}
 \centering
 \includegraphics[keepaspectratio=true,width=0.48\textwidth]{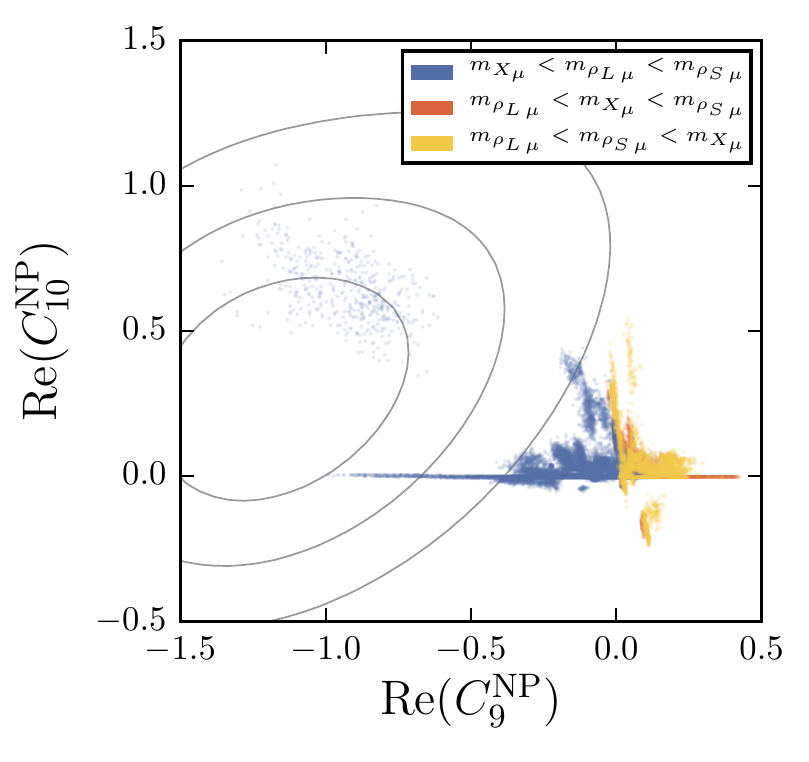}
 \caption{Predictions for the Wilson coefficients $C_{9}^{bs}$ and $C_{10}^{bs}$ for
 the vector resonance $X_\mu$ beeing the lightest one (blue points), having a mass between the other four light and three heavy resonances (orange points) and for the case where $X_\mu$ is the heaviest resonance (yellow points).
 We also show contour lines for the boundaries of the one, two and three sigma regions around the best fit value calculated by \texttt{flavio}~\cite{flavio}.
 } \label{fig:C9_C10}
\end{figure}

Similarly to the minimal coset, we find potentially sizable contributions to the Wilson
coefficients $C_{9,10}$ of the semi-leptonic operators contributing to rare semi-leptonic
${b\to s\,\ell^+\ell^-}$ transitions,
\begin{align}
O_9 &= (\bar s_L \gamma^\mu b_L) (\bar \ell \gamma_\mu \ell)
\,,&
O_{10} &= (\bar s_L \gamma^\mu b_L) (\bar \ell \gamma_\mu\gamma_5 \ell)
\,.&
\end{align}
We again find that large effects typically come from two types of contributions
(see \cite[Section 3.2.5]{Niehoff:2015iaa} for a detailed discussion),
\begin{itemize}
\item a ``KK $Z$-like'' contribution contributing mostly to $C_{10}$, almost
always destructively interfering with the SM contribution, leading to a suppression
of the $B_s\to\mu^+\mu^-$ branching ratio, or
\item a ``KK photon-like'' contribution contributing only to $C_9$ and almost
always destructively interfering with the SM. We find
a NP effect up to $C_9\approx -1$, which is mediated by a vector resonance that
is mostly $X_\mu$, with a mass around 1~TeV (not constrained by the $S$
parameter at tree-level) and a dominant decay to top quark pairs (cf. section \ref{sec:bounds_vectors}).
\end{itemize}
Interestingly, the case of negative NP contribution to $C_9$ could explain
various tensions observed in global fits to $b\to s\,\mu^+\mu^-$ transitions,
in particular in the angular distribution of $B\to K^*\mu^+\mu^-$ and in the
branching ratios of $B\to K\mu^+\mu^-$ and $B_s\to \phi\,\mu^+\mu^-$
(see e.g.\ \cite{Altmannshofer:2014rta,Straub:2015ica,Descotes-Genon:2015uva,Hurth:2016fbr}).
Figure~\ref{fig:C9_C10} shows all viable parameter points in the plane of
new physics contributions to $C_9$ and $C_{10}$ compared to a the results of
a global fit to $b \to s\,\mu^+\mu^-$ and $b\to s\gamma$ processes performed with
\texttt{flavio}~\cite{flavio} as an update
of \cite{Altmannshofer:2014rta}. The colour coding of the points
gives an indication of the mass hierarchies in the spin-1 sector, demonstrating
that for points with large negative contribution to $C_9$, the lightest spin-1
resonance is always dominantly $X_\mu$.

We note that, since
we assume elementary leptons however, our model cannot explain the apparent violation
of lepton flavour universality in $B\to K\mu^+\mu^-$ (see \cite{Niehoff:2015bfa}
for an attempt to explain it with partially composite leptons).

\subsubsection{Neutron electric dipole moment}

In contrast to the minimal coset studied in \cite{Niehoff:2015iaa}, where the
neutron EDM was found to be well below current bounds, due to the presence
of spontaneous $C\! P$ violation in the scalar sector we now find appreciable
contributions induced at the one-loop level from $h$ and $\eta$ exchange.
We have imposed this constraint on our otherwise allowed points \textit{a posteriori}
and found that roughly one third of the points were excluded by the neutron EDM,
treated as discussed in section~\ref{sec:constraints}.
We conclude that a neutron EDM in the ball park of near-future experiments is
a generic prediction of the model and provides a clear distinction from the
model with minimal coset studied in \cite{Niehoff:2015iaa}.

\section{Conclusions} \label{sec:conclusion}
In this work, we have performed a comprehensive numerical analysis of a
composite pNGB Higgs model, based on the symmetry breaking coset $\text{SO}(6)/\text{SO}(5)$,
with partial compositeness and a flavour symmetry
in the fermion sector to protect flavour-changing neutral currents.
Compared to the minimal custodial coset $\text{SO}(5)/\text{SO}(4)$ (analyzed recently by us in~\cite{Niehoff:2015iaa}), the most striking feature of the enlarged coset is the presence of an additional pseudoscalar pNGB degree of freedom that can mix with the Higgs boson.
We investigated the dynamics of this scenario, placing
special emphasis on realistic electroweak symmetry breaking that reproduces the correct Higgs mass and vacuum expectation value, as well as on the collider phenomenolgy of
the additional scalar $\eta$.

Our main findings are summarized as follows.
\begin{itemize}
 \item A minimum for the effective potential is found for all reasonable vacuum expectation values of the two scalars $\gb{h}$ and $\gb{\eta}$ that yield the correct Fermi constant (cf. left plot in figure~\ref{fig:sh_vs_setatilde}).
       Therefore, direct and indirect bounds do not constrain the potential in this respect.
 \item For a symmetry breaking scale $f < 1 \, \text{TeV}$, a Barbieri-Giudice measure for fine-tuning well below $100$ is possible, showing no immediate tension with fine-tuning for this model (see the right plot in figure~\ref{fig:sh_vs_setatilde}).
 \item The pseudoscalar $\gb{\eta}$ can mix with the Higgs if $C\! P$ is broken by the effective potential.
       However, we find this not to be in conflict with measurements of the Higgs signal strengths (see figure~\ref{fig:higgs_signalstrengths}).
       This is mainly due to relatively large uncertainties for these observables, so a precision measurement of the Higgs couplings can potentially cut deeply into the parameter space.
 \item Values for the mass of the scalar $\eta$ are found over the large range $130-1600$~GeV (we do not allow for $\eta$ to be lighter than the SM-like Higgs).
 \item The production of $\eta$ is clearly dominated by gluon fusion and suppressed compared to Higgs production mainly due to its higher mass and the therefore smaller gluon parton luminosity.
 \item If kinematically allowed, the highest branching ratios for the decay of $\eta$ are found for two Higgses in the final state, followed by $t\bar{t}$ and a pair of $W$ or $Z$ bosons. The by far strongest experimental bounds can be set in the two Higgs channel.
 While the constraints in $t\bar{t}$ are weak, the higher experimental sensitivity in the diboson channels makes them also promising for setting bounds. The branching ratio to diphotons is usually tiny and no signal is expected in this channel in the near future.
 As we find a mass for $\eta$ below 1~TeV for most of the parameter points, we again want to stress the importance of experimental searches in the promising channels also
 for sub-TeV masses.
\item The bounds on neutral electroweak resonances depend on their mass spectrum.
For the case of $X_\mu$ being the lightest vector resonance, its decays in the dilepton and $t\bar{t}$ channels yield the strongest bounds.
For $\rho_{L\,\mu}$ and $\rho_{R\,\mu}$ being lighter than $X_\mu$, the strongest constraints arise from the decays in the dilepton and diboson channels.
 \item While the charged vector resonances are slightly less constrained than the neutral ones, still viable parameter points are expected to be probed by the $\rho_{L\,\mu}^\pm$ state in the diboson and $\ell^\pm\nu$ channels in the near future.
 \item We find recently published searches for fermion resonances to strongly constrain the parameter space.
 Soon possible experimental analyses of quark resonance pair production that are sensitive to masses of more than $1.2$~TeV are expected to probe nearly all of the viable parameter points found in our scan.
 However, especially the decays with light SM quarks in the final state are very poorly covered by current LHC searches.
 \item Meson-antimeson mixing observables are strong constraints on the model,
 inspite of the large flavour symmetry. Given theoretical uncertainties, the
 most promising observable to observe deviations from the SM
 is mixing-induced CP violation in the $B_s$ system
 that can saturate present experimental bounds that will be strongly improved
 in the near future.
 \item Rare $B$ decays based on the $b\to s\ell^+\ell^-$ transition, such as
 $B_s\to\mu^+\mu^-$ or $B\to K^{(*)}\mu^+\mu^-$, can also be affected. The branching
 ratios are predicted to be suppressed with respect to the SM for almost allow
 viable points. Currently observed tensions with the SM expectations could be explained
 and would imply the presence of a vector resonance with a mass of about 1~TeV
 decaying dominantly to $t \bar t$.
 \item The electric dipole moment of the neutron can reach values in the ball park
 of the current experimental bound and could be in reach of near-future searches.
 This prediction distinguishes the model at hand from the model with the minimal coset
 $\text{SO}(5)/\text{SO}(4)$ and the flavour symmetry $\text{U}(2)^3_\text{RC}$ studied in
 \cite{Niehoff:2015iaa}
 and is due to the additional source of $C\! P$ violation present in the scalar sector.
\end{itemize}

In summary, we find this next-to-minimal composite pNGB Higgs model to pass
both the direct and the indirect experimental constraints considered by us with
a moderate electroweak fine-tuning, while prospects to discover or further constrain
the model are promising at run 2 of the LHC and in future precision-frontier experiments.

\subsection*{Acknowledgements}
It is a pleasure to thank
Rikkert Frederix,
Denis Karateev and
Andreas Trautner
for helpful discussions.
We are grateful for the support by Jovan Mitrevski through the Computational Center for Particle and Astrophysics (C2PAP), where the simulations have been carried out.
PS would like to thank CP$^3$-Origins for the hospitality during the final stages of this work.
CN thanks the University of Cincinnati for the hospitality during finalizing this work.
This work was supported by the DFG cluster of excellence ``Origin and Structure of the Universe''.

\appendix
\section{$\text{SO}(6)$} \label{appendix:SO6}
\subsection{Generators} \label{appendix:SO6_generators}
The group theory of $\text{SO}(6)$ is very nicely explained in~\cite{Mrazek:2011iu}.
$\text{SO}(6)$ has 15 generators: the usual 10 generators of $\text{SO}(5)$ ($\T^a_L$, $\T^a_R$, $\T^i_{\hat 1}$) plus a bidoublet ($\T^i_{\hat 2}$) plus a singlet ($\T_S$).
The breaking of the latter two gives five NGBs containing the Higgs doublet $\Phi$ and the additional pseudoscalar $\eta$
\footnote{Here, we use a slightly different convention compared to~\cite{Mrazek:2011iu} to match the convention used in the $\text{SO}(5) / \text{SO}(4)$ models.},
\begin{align*}
 \left[ \T^a_L \right]_{I J} &= -\frac{\im}{2} \left[ \frac{1}{2} \epsilon^{abc} \left( \delta_{b I} \delta_{c J} - \delta_{b J} \delta_{c I} \right) + \left( \delta_{a I} \delta_{4 J} - \delta_{a J} \delta_{4 I} \right) \right] \\
 \left[ \T^a_R \right]_{I J} &= -\frac{\im}{2} \left[ \frac{1}{2} \epsilon^{abc} \left( \delta_{b I} \delta_{c J} - \delta_{b J} \delta_{c I} \right) - \left( \delta_{a I} \delta_{4 J} - \delta_{a J} \delta_{4 I} \right) \right] \\
 \left[ \T^i_{\hat 1} \right]_{I J} &=  - \frac{\im}{\sqrt{2}} \left( \delta_{i I} \delta_{5 J} - \delta_{i J} \delta_{5 I} \right) \\
 \left[ \T^i_{\hat 2} \right]_{I J} &=  - \frac{\im}{\sqrt{2}} \left( \delta_{i I} \delta_{6 J} - \delta_{i J} \delta_{6 I} \right) \\
 \left[ \T_S \right]_{I J} &=  - \frac{\im}{\sqrt{2}} \left( \delta_{5 I} \delta_{6 J} - \delta_{5 J} \delta_{6 I} \right),
\end{align*}
where $a \in \{1,2,3\}$, $i \in \{1,2,3,4\}$ and $I,J \in \{1,2,3,4,5,6\}$.

\subsection{Elementary embeddings} \label{appendix:SO6_embeddings}
The incomplete embeddings of elementary quarks into fundamentals of  $\text{SO}(6)$, as used in (\ref{eq:CompElemMixings}), are given as
\begin{align*}
 \xi_{u\text{L}}   &= \frac{1}{\sqrt{2}} \left( \begin{array}{c} d_\text{L}\\-\im d_\text{L}\\u_\text{L}\\ \im u_\text{L}\\0\\0 \end{array} \right), &
 \xi^5_{u\text{R}} &=                    \left( \begin{array}{c}  0\\0\\0\\0\\u_\text{R}\\0\end{array} \right), &
 \xi^6_{u\text{R}} &=                    \left( \begin{array}{c}  0\\0\\0\\0\\0\\u_\text{R}\end{array} \right), \\
 \xi_{d\text{L}}   &= \frac{1}{\sqrt{2}} \left( \begin{array}{c} u_\text{L}\\ \im u_\text{L}\\-d_\text{L}\\ \im u_\text{L}\\0\\0 \end{array} \right), &
 \xi^5_{d\text{R}} &=                    \left( \begin{array}{c}  0\\0\\0\\0\\d_\text{R}\\0\end{array} \right), &
 \xi^6_{d\text{R}} &=                    \left( \begin{array}{c}  0\\0\\0\\0\\0\\d_\text{R}\end{array} \right). \\
\end{align*}

\section{Mass matrices} \label{appendix:MassMatrices}
In this section we explicitly give the mass matrices of bosons and fermions for the model considered in this work.
In the way they are presented here, the mass matrices depend on the scalar fields $\gb{h}$ and $\gb{\eta}$.
By this, they are suitable for the calculation of the effective potential via eq. (\ref{eq:EffPotGeneral}), which determines the vevs of the scalar fields.
In the end, the physical masses of the particles are given as the eigenvalues of the mass matrices, where the scalar fields take their vacuum value, $M_i(\gb{h}=v_h, \gb{\eta}=v_\eta)$.

\subsection{Vector bosons} \label{appendix:MassMatrices:Vector}
In the vector boson sector it is convenient to group the fields by their charge, such that there are separate mass matrices for neutral and charged vector bosons.
For the neutral fields the elementary $W_\mu^3$- and $B_\mu$-bosons mix with composite vectors via composite-elementary mixings, $v^0_W$ and $v^0_B$.
The structure of the mass matrix is the following
\begin{equation}
 M^2_Z(\gb{h}, \gb{\eta}) =
 \left( \begin{array}{c||c|c|c|c|c|c|c|c|c|c}
                      &  W_{\mu}^3 & B_\mu & \rho_{L \, \mu}^3 & \rho_{R \, \mu}^3 & \axial_{1 \, \mu}^3 & \axial_{2 \, \mu}^3 & X_\mu & \axial_{1 \, \mu}^4 & \axial_{2 \, \mu}^4 & \rho_{S \, \mu} \\
  \hline \hline
  W_{\mu}^3           & \multicolumn{2}{c|}{\multirow{2}{*}{$D^0_1$}}          & \multicolumn{5}{c|}{v_W^{0 \, \text{t}}}      & \multicolumn{3}{c}{\multirow{7}{*}{0}} \\
  \cline{1-1} \cline{4-8}
  B_\mu               & \multicolumn{2}{c|}{}                                  & \multicolumn{5}{c|}{v_B^{0 \, \text{t}}}      & \multicolumn{3}{c}{} \\
  \cline{1-8}
  \rho_{L \, \mu}^3   & \multirow{5}{*}{$v_W^0$}   & \multirow{5}{*}{$v_B^0$}  & \multicolumn{5}{c|}{\multirow{5}{*}{$D^0_2$}} & \multicolumn{3}{c}{}\\
  \cline{1-1}
  \rho_{R \, \mu}^3   &                            &                           & \multicolumn{5}{c|}{}                         & \multicolumn{3}{c}{}\\
  \cline{1-1}
  \axial_{1 \, \mu}^3 &                            &                           & \multicolumn{5}{c|}{}                         & \multicolumn{3}{c}{}\\
  \cline{1-1}
  \axial_{2 \, \mu}^3 &                            &                           & \multicolumn{5}{c|}{}                         & \multicolumn{3}{c}{}\\
  \cline{1-1}
  X_\mu               &                            &                           & \multicolumn{5}{c|}{}                         & \multicolumn{3}{c}{}\\
  \cline{1-11}
  \axial_{1 \, \mu}^4 & \multicolumn{7}{c|}{\multirow{3}{*}{0}}                                                                & \multicolumn{3}{c}{\multirow{3}{*}{$D^0_3$}} \\
  \cline{1-1}
  \axial_{2 \, \mu}^4 & \multicolumn{7}{c|}{}                                                                                  &  \multicolumn{3}{c}{}\\
  \cline{1-1}
  \rho_{S \, \mu}     & \multicolumn{7}{c|}{}                                                                                  &  \multicolumn{3}{c}{}
 \end{array} \right),
\end{equation}
where the diagonal elements are given by
\begin{subequations}
\begin{align}
 D^0_1 &=  \left( \begin{array}{c||cc}           &  W_{\mu}^3            & B_\mu    \\
                                       \hline \hline
                                       W_{\mu}^3 & \frac{f_1^2 g_0^2}{2} &                                                   \\
                                       B_\mu     &                       & \frac{1}{2} \left(f_1^2+f_X^2\right) g_0^{' \, 2} \end{array} \right), \\
 D^0_2 &= \left( \begin{array}{c||cccccc}                     & \rho_{L \, \mu}^3        & \rho_{R \, \mu}^3        & \axial_{1 \, \mu}^3      & \axial_{2 \, \mu}^3                             & X_\mu \\
                                                              \hline \hline
                                          \rho_{L \, \mu}^3   & \frac{f_1^2 g_\rho^2}{2} &                          &                          &                                                 &       \\
                                          \rho_{R \, \mu}^3   &                          & \frac{f_1^2 g_\rho^2}{2} &                          &                                                 &       \\
                                          \axial_{1 \, \mu}^3 &                          &                          & \frac{f_1^2 g_\rho^2}{2} &                                                 &       \\
                                          \axial_{2 \, \mu}^3 &                          &                          &                          & \frac{f_1^4 g_\rho^2}{2 \left(f_1^2-f^2\right)} &       \\
                                          X_\mu               &                          &                          &                          &                                                 & \frac{f_X^2 g_X^2}{2} \end{array} \right), \\
 D^0_3 &= \left( \begin{array}{c||ccc}                     & \axial_{1 \, \mu}^4 & \axial_{2 \, \mu}^4 & \rho_{S \, \mu} \\
                                                           \hline\hline
                                       \axial_{1 \, \mu}^4 & \frac{f_1^2 g_\rho^2}{2} &                                                 &                                                 \\
                                       \axial_{2 \, \mu}^4 &                          & \frac{f_1^4 g_\rho^2}{2 \left(f_1^2-f^2\right)} &                                                 \\
                                       \rho_{S \, \mu}     &                          &                                                 & \frac{f_1^4 g_\rho^2}{2 \left(f_1^2-f^2\right)} \end{array} \right)
\end{align}
\end{subequations}
and the ($\gb{h}$- and $\gb{\eta}$-dependent) composite-elementary mixings are
\begin{subequations}
\begin{align}
 v_W^0 &= \left( \begin{array}{c||c}                     &  W_{\mu}^3                                                                       \\
                                                         \hline\hline
                                     \rho_{L \, \mu}^3   & -\frac{1}{4} f_1^2 \, g_0 g_\rho \, \left(c_h \cetatilde^2+\setatilde^2+1\right) \\
                                     \rho_{R \, \mu}^3   & -\frac{1}{4} f_1^2 \, g_0 g_\rho \, \left(1 - c_h\right) \cetatilde^2            \\
                                     \axial_{1 \, \mu}^3 & \frac{f_1^2 \, g_0 g_\rho \, (1-c_h) \setatilde \cetatilde}{2 \sqrt{2}}          \\
                                     \axial_{2 \, \mu}^3 & -\frac{f_1^2 \, g_0 g_\rho \,  \sh \cetatilde}{2 \sqrt{2}}                       \\
                                     X_\mu               & 0                                                                                \end{array} \right), \\
 v_B^0 &= \left( \begin{array}{c||c}                     &  B_{\mu}                                                                             \\
                                                         \hline\hline
                                     \rho_{L \, \mu}^3   &  -\frac{1}{4} f_1^2 \, g_0^{'} g_\rho \, (1-c_h) \cetatilde^2                        \\
                                     \rho_{R \, \mu}^3   & -\frac{1}{4} f_1^2 \, g_0^{'} g_\rho \, \left(c_h \cetatilde^2+\setatilde^2+1\right) \\
                                     \axial_{1 \, \mu}^3 & -\frac{f_1^2 \, g_0^{'} g_\rho \, (1-c_h) \setatilde \cetatilde}{2 \sqrt{2}}         \\
                                     \axial_{2 \, \mu}^3 & \frac{f_1^2 \, g_0^{'} g_\rho \, \sh \cetatilde}{2 \sqrt{2}}                         \\
                                     X_\mu               & -\frac{1}{2} f_X^2 \, g_0^{'} g_X                                                    \end{array} \right).
\end{align}
\end{subequations}
After the scalar fields take their vevs the neutral boson mass matrices will have one massless and one rather light (as compared to the scale $f$) eigenvalue.
These we will identify with the photon and the $Z$-boson, respectively.

The mass matrix of the charged vector bosons takes a similar (but easier) form,
\begin{equation}
  M^2_W(\gb{h}, \gb{\eta}) = \left(
\begin{array}{c||c|c|c|c|c}
                     &  W^+_\mu & \rho^+_{L \, \mu} & \rho^+_{R \, \mu} & \axial^+_{1 \, \mu} & \axial^+_{2 \, \mu} \\
                     \hline\hline
 W^-_\mu             & D^+_1                    & \multicolumn{4}{c}{v^{+ \, \text{t}}_W}      \\
 \hline
 \rho^-_{L \, \mu}   & \multirow{4}{*}{$v^+_W$} & \multicolumn{4}{c}{\multirow{4}{*}{$D^+_2$}} \\
 \cline{1-1}
 \rho^-_{R \, \mu}   &                          & \multicolumn{4}{c}{}                         \\
 \cline{1-1}
 \axial^-_{1 \, \mu} &                          & \multicolumn{4}{c}{}                         \\
 \cline{1-1}
 \axial^-_{2 \, \mu} &                          & \multicolumn{4}{c}{}                         \end{array} \right).
\end{equation}
Here, the charged vector bosons $V^\pm_\mu$ are given as linear combinations
\begin{equation}
 V^\pm_\mu = \frac{1}{\sqrt{2}} \left( V^1_\mu \mp \im V^2_\mu \right),
\end{equation}
were the $1$ and $2$ refer to the $\text{SU}(2)$ indices of the vector triplets.
The mass matrix has diagonal elements
\begin{subequations}
\begin{align}
 D^+_1 &= \left( \begin{array}{c||c}
                                               & W^+_\mu             \\
                                               \hline\hline
                                     W^-_\mu & \frac{f_1^2 g_0^2}{2} \end{array} \right), \\
 D^+_1 &= \left( \begin{array}{c||cccc}
                                                            & \rho^+_{L \, \mu}  & \rho^+_{R \, \mu} & \axial^+_{1 \, \mu} & \axial^+_{2 \, \mu} \\
                                                            \hline \hline
                                        \rho^-_{L \, \mu}   & \frac{f_1^2 g_\rho^2}{2} &                          &                          &                                                 \\
                                        \rho^-_{R \, \mu}   &                          & \frac{f_1^2 g_\rho^2}{2} &                          &                                                 \\
                                        \axial^-_{1 \, \mu} &                          &                          & \frac{f_1^2 g_\rho^2}{2} &                                                 \\
                                        \axial^-_{2 \, \mu} &                          &                          &                          & \frac{f_1^4 g_\rho^2}{2 \left(f_1^2-f^2\right)} \end{array} \right),
\end{align}
\end{subequations}
and composite-elementary mixings
\begin{align}
 v^+_W = \left( \begin{array}{c||c}
                                                        &  W^+_\mu                                                                                                               \\
                                                        \hline \hline
                                    \rho^-_{L \, \mu}   & -\frac{1}{4} f_1^2 \, g_0 g_\rho \, \left(c_h \cetatilde^2+\setatilde^2+1\right) \\
                                    \rho^-_{R \, \mu}   & -\frac{1}{4} f_1^2 \, g_0 g_\rho \, (1-c_h) \cetatilde^2       \\
                                    \axial^-_{1 \, \mu} &  \frac{f_1^2 \, g_0 g_\rho \, (1-c_h) \setatilde \cetatilde}{2 \sqrt{2}} \\
                                    \axial^-_{2 \, \mu} & -\frac{f_1^2 \, g_0 g_\rho \, \sh \cetatilde}{2 \sqrt{2}}      \end{array} \right).
\end{align}
Also this mass matrix has a light eigenvalue which is the $W$-boson of the SM.

The mass matrices for gluons and their resonances can be found in~\cite[Appendix B.1]{Niehoff:2015iaa}.

\subsection{Fermions} \label{appendix:MassMatrices:Fermion}
As for the vector bosons the fermion mass matrices are also best grouped by the electrical charge of the fields, i.e. there is a mass matrix for up-type as well as for down-type quarks and quark resonances.
The mass matrices for heavy resonances with exotic charges $q=+ \frac{5}{3}$ and $q=-\frac{4}{3}$ are given in~\cite[Appendix B.2]{Niehoff:2015iaa}.

The $\gb{h},\gb{\eta}$-dependent mass matrix for up-type quarks is given as
\begin{equation}
 M_u(\gb{h}, \gb{\eta}) = \left(
\begin{array}{c||c|cc|cc|cc|cc|cc}
   &  u_\text{R} & Q_{u \text{R}}^{+-} & \widetilde{Q}_{u \text{R}}^{+-} & Q_{u \text{R}}^{-+} & \widetilde{Q}_{u \text{R}}^{-+} & Q_{d \text{R}}^{++} & \widetilde{Q}_{d \text{R}}^{++} & S^1_{u \text{R}} & \widetilde{S}^1_{u \text{R}} & S^2_{u \text{R}} & \widetilde{S}^2_{u \text{R}} \\
   \hline \hline
   \overline{u}_\text{L}                      & 0 & \multicolumn{2}{c|}{\underline{\Delta}_{Q u \text{L}}^{+-}}
                                                  & \multicolumn{2}{c|}{\underline{\Delta}_{Q u \text{L}}^{-+}}
                                                  & \multicolumn{2}{c|}{\underline{\Delta}_{Q d \text{L}}^{++}}
                                                  & \multicolumn{2}{c|}{\underline{\Delta}_{S u \text{L}}^{1}}
                                                  & \multicolumn{2}{c }{\underline{\Delta}_{S u \text{L}}^{2}} \\
   \hline
   \overline{Q}_{u \text{L}}^{+-}             & \multirow{2}{*}{$\underline{\Delta}_{Qu\text{R}}^{+- \, \dagger}$} & \multicolumn{2}{c|}{\multirow{2}{*}{$\underline{M}_{u}$}} & & & & & & &\\
   \overline{\widetilde{Q}}_{u \text{L}}^{+-} &  & \multicolumn{2}{c|}{\multirow{2}{*}{}} & & & & & & &\\
   \hline
   \overline{Q}_{u \text{L}}^{-+}             & \multirow{2}{*}{$\underline{\Delta}_{Qu\text{R}}^{-+ \, \dagger}$} & & & \multicolumn{2}{c|}{\multirow{2}{*}{$\underline{M}_{u}$}} & & & & &\\
   \overline{\widetilde{Q}}_{u \text{L}}^{-+} &  & & & \multicolumn{2}{c|}{\multirow{2}{*}{}} & & & & &\\
   \hline
   \overline{Q}_{d \text{L}}^{++}             & \multirow{2}{*}{$0$} & & & & & \multicolumn{2}{c|}{\multirow{2}{*}{$\underline{M}_{d}$}} & & &\\
   \overline{\widetilde{Q}}_{d \text{L}}^{++} & & & & & & \multicolumn{2}{c|}{\multirow{2}{*}{}} & & &\\
   \hline
   \overline{S}^1_{u \text{L}}                & \multirow{2}{*}{$\underline{\Delta}_{Su\text{R}}^{1 \, \dagger}$} & & & & & & & \multicolumn{2}{c|}{\multirow{2}{*}{$\underline{M}_{u}$}} &\\
   \overline{\widetilde{S}}^1_{u \text{L}}    & & & & & & & & \multicolumn{2}{c|}{\multirow{2}{*}{}} & &\\
   \hline
   \overline{S}^2_{u \text{L}}                & \multirow{2}{*}{$\underline{\Delta}_{Su\text{R}}^{2 \, \dagger}$} & & & & & & & & &\multicolumn{2}{c}{\multirow{2}{*}{$\widehat{\underline{M}}_{u}$}}\\
   \overline{\widetilde{S}}^2_{u \text{L}}    & & & & & & & & & &\multicolumn{2}{c}{\multirow{2}{*}{}}\\
 \end{array}\right). \label{eq:MassMatrix:U}
\end{equation}
By construction, the left-handed elementary quarks mix with heavy composite states $\Psi$ while the right-handed ones mix with composites $\widetilde \Psi$ (see eq. (\ref{eq:CompElemMixings})).
Both composites mix via non-diagonal composite mass matrices
\begin{subequations}
\begin{align}
 \underline{M}_u &= \left( \begin{array}{c||cc}
                                                                                         & \Psi_{u \text{R}} & \widetilde{\Psi}_{u \text{R}} \\
                                                                                         \hline \hline
                                                \overline{\Psi}_{u_\text{L}}             & m_U & m_{Y u}\\
                                                \overline{\widetilde{\Psi}}_{u \text{L}} &   0 & m_{\widetilde{U}}
                           \end{array} \right), \\
 \underline{M}_d &= \left( \begin{array}{c||cc}
                                                                                         & \Psi_{d \text{R}} & \widetilde{\Psi}_{d \text{R}} \\
                                                                                         \hline \hline
                                                \overline{\Psi}_{d_\text{L}}             & m_D & m_{Y d}\\
                                                \overline{\widetilde{\Psi}}_{d \text{L}} &   0 & m_{\widetilde{D}}
                           \end{array} \right), \\
 \widehat{\underline{M}}_u &= \left( \begin{array}{c||cc}
                                                                                         & \Psi_{u \text{R}} & \widetilde{\Psi}_{u \text{R}} \\
                                                                                         \hline \hline
                                                \overline{\Psi}_{u_\text{L}}             & m_U & m_{Y u} + Y_u\\
                                                \overline{\widetilde{\Psi}}_{u \text{L}} &   0 & m_{\widetilde{U}}
                           \end{array} \right), \\
 \widehat{\underline{M}}_d &= \left( \begin{array}{c||cc}
                                                                                         & \Psi_{d \text{R}} & \widetilde{\Psi}_{d \text{R}} \\
                                                                                         \hline \hline
                                                \overline{\Psi}_{d_\text{L}}             & m_D & m_{Y d} + Y_d\\
                                                \overline{\widetilde{\Psi}}_{d \text{L}} &   0 & m_{\widetilde{D}}
                           \end{array} \right).
\end{align}
\end{subequations}
The composite-elementary mixings carry the $\gb{h}, \gb{\eta}$-dependence.
These are given as
\begin{subequations}
\begin{align}
 \underline{\Delta}_{Q u \text{L}}^{+-} &= \left( \begin{array}{c||cc}
                                                                                             & Q_{u \text{R}}^{+-} & \widetilde{Q}_{u \text{R}}^{+-} \\
                                                                                             \hline\hline
                                                                       \overline{u}_\text{L} & -\frac{1}{2} \Delta_{u \mathrm{L}} \left(c_h \cetatilde^2+1\right)  &  0 \end{array} \right), \\
 \underline{\Delta}_{Q u \text{L}}^{-+} &= \left( \begin{array}{c||cc}
                                                                                             & Q_{u \text{R}}^{-+} & \widetilde{Q}_{u \text{R}}^{-+} \\
                                                                                             \hline\hline
                                                                       \overline{u}_\text{L} & \frac{1}{2} \Delta_{u \mathrm{L}} \cetatilde^2 \left(1 - c_h \right)  &  0 \end{array} \right), \\
 \underline{\Delta}_{Q d \text{L}}^{++} &= \left( \begin{array}{c||cc}
                                                                                             & Q_{d \text{R}}^{++} & \widetilde{Q}_{d \text{R}}^{++} \\
                                                                                             \hline\hline
                                                                       \overline{u}_\text{L} & -\Delta_{d \text{L}}  &  0 \end{array} \right), \\
 \underline{\Delta}_{S u \text{L}}^{1}  &= \left( \begin{array}{c||cc}
                                                                                             & S_{u \text{R}}^{1} & \widetilde{S}_{u \text{R}}^{1} \\
                                                                                             \hline\hline
                                                                       \overline{u}_\text{L} & -\frac{\im}{\sqrt{2}} \Delta_{u \mathrm{L}} \left( 1-c_h \right) \setatilde \cetatilde  &  0 \end{array} \right), \\
 \underline{\Delta}_{S u \text{L}}^{2}  &= \left( \begin{array}{c||cc}
                                                                                             & S_{u \text{R}}^{2} & \widetilde{S}_{u \text{R}}^{2} \\
                                                                                             \hline\hline
                                                                       \overline{u}_\text{L} & \frac{\im}{\sqrt{2}} \Delta_{u \mathrm{L}} \cetatilde \sh  &  0 \end{array} \right), \\
 \underline{\Delta}_{Q u \text{R}}^{+-} &= \left( \begin{array}{c||cc}
                                                                                             & \overline{Q}_{u \text{L}}^{+-} & \widetilde{\overline{Q}}_{u \text{L}}^{+-} \\
                                                                                             \hline\hline
                                                                                  u_\text{R} & 0  &  \frac{\im}{\sqrt{2}} \left(\Delta_{u \mathrm{R}}^{5} \left( (1-c_h) \setatilde \cetatilde \right)+\Delta_{u \mathrm{R}}^{6} \sh \cetatilde\right) \end{array} \right),\\
 \underline{\Delta}_{Q u \text{R}}^{-+} &= \left( \begin{array}{c||cc}
                                                                                             & \overline{Q}_{u \text{L}}^{-+} & \widetilde{\overline{Q}}_{u \text{L}}^{-+} \\
                                                                                             \hline\hline
                                                                                  u_\text{R} & 0  &  \frac{\im}{\sqrt{2}} \left(\Delta_{u \mathrm{R}}^{5} \left( (1-c_h) \setatilde \cetatilde \right)+\Delta_{u \mathrm{R}}^{6} \sh \cetatilde\right) \end{array} \right),\\
 \underline{\Delta}_{S u \text{R}}^{1}  &= \left( \begin{array}{c||cc}
                                                                                             & \overline{S}_{u \text{L}}^{1} & \widetilde{\overline{S}}_{u \text{L}}^{1} \\
                                                                                             \hline\hline
                                                                                  u_\text{R} & 0  &  - \Delta_{u \mathrm{R}}^{5} \left(\cetatilde^2+c_h \setatilde^2\right) + \Delta_{u \mathrm{R}}^{6} \sh \setatilde  \end{array} \right),\\
 \underline{\Delta}_{S u \text{R}}^{2}  &= \left( \begin{array}{c||cc}
                                                                                             & \overline{S}_{u \text{L}}^{2} & \widetilde{\overline{S}}_{u \text{L}}^{2} \\
                                                                                             \hline\hline
										u_\text{R} & 0  &  - \Delta_{u \mathrm{R}}^{5} \sh \setatilde - \Delta_{u \mathrm{R}}^{6} c_h  \end{array} \right).
\end{align}
\end{subequations}
The mass matrix for down-type states takes an analogue form of (\ref{eq:MassMatrix:U}).
One can get it by replacing $u \leftrightarrow d$ and $+ \leftrightarrow -$ (for the $\text{SU}(2)_\text{L} \times \text{SU}{2}_\text{R}$ indices) in the up-type mass matrix.

\section{Field shift in the boson sector} \label{appendix:BosonFieldshift}
In holographic gauge the Lagrangian (\ref{eq:BosonLagrangian}) leads to kinematic mixings between the pNGB and the composite vector bosons.
These have to be canceled by field shifts and rescalings of the boson fields~\cite{Callan:1969sn, Parganlija:2012xj}\footnote{For the decay constants $f, f_1$ in the composite sector we use the notation of~\cite {DeCurtis:2011yx}.
          In particular, $f$ is defined through $1/f^2 = 1/f_1^2 + 1/f_2^2$.},
\begin{align}
 \axial^1_{4 \, \mu} &\rightarrow  \axial^1_{4 \, \mu} +  \frac{\left(\sqrt{2} \cos \left(\frac{v_h}{f}\right)-\sqrt{2}\right)}{f_1 g_\rho} \partial_\mu \gb{\eta},\\
 \axial^2_{4 \, \mu} &\rightarrow  \axial^2_{4 \, \mu} + \frac{\sqrt{2} \left(f^2-f_1^2\right) \cos \left(\frac{v_\eta}{f \sin\left(  \frac{v_h}{f} \right)}\right)}{f_1^3 g_\rho} \partial_\mu \gb{h}  -\frac{\sqrt{2} \left(f^2-f_1^2\right) \sin \left(\frac{v_h}{f}\right) \sin \left(\frac{v_\eta}{f \sin\left(  \frac{v_h}{f} \right)}\right)}{f_1^3 g_\rho}  \partial_\mu \gb{\eta},\\
 \rho_{S \, \mu}     &\rightarrow  \rho_{S \, \mu}     +\frac{\sqrt{2} \left(f^2-f_1^2\right) \sin \left(\frac{v_\eta}{f \sin\left(  \frac{v_h}{f} \right)}\right)}{f_1^3 g_\rho}  \partial_\mu \gb{h}  +\frac{\sqrt{2} \left(f^2-f_1^2\right) \sin \left(\frac{v_h}{f}\right) \cos \left(\frac{v_\eta}{f \sin\left(  \frac{v_h}{f} \right)}\right)}{f_1^3 g_\rho}  \partial_\mu \gb{\eta}	,
\end{align}
\begin{equation} \label{eq:ScalarRot}
 \gb{h} \rightarrow \frac{f_1}{f} \gb{h}, \qquad \gb{\eta} \rightarrow \frac{f_1}{f \sin\left( \frac{v_h}{f} \right)} \gb{\eta}.
\end{equation}
where $v_h$ and $v_\eta$ denote the vevs of $\gb{h}$ and $\gb{\eta}$, respectively.
As a consequence,
all scalar interaction terms only depend on $\sh$ and $\setatilde$ (cf. equation (\ref{eq:sh_setatilde_def})).

\section{Flavour structure of the composite-elementary mixings} \label{appendix:FlavourStructure}
We assume the composite sector to be invariant under an $\text{U}(2)^3$ flavour symmetry that is only broken by the couplings of the left-handed elementary quarks to the composite sector.
For this one assumes that the first two generations form a doublet under this symmetry while the third one is regarded as a singlet.
Using the notations of~\cite{Barbieri:2012uh} we parametrize the composite-elementary mixings in a spurion expansion as follows
\begin{subequations}
\begin{align}
\Delta_{u \text{L}} &=
\begin{pmatrix}
c_u\, \Delta_{u_1\text{L}} & -s_u\, \Delta_{u_2\text{L}}\, e^{\im\alpha_u} \\
s_u\, \Delta_{u_1\text{L}}\, e^{-\im\alpha_u} & c_u\, \Delta_{u_2\text{L}} &
\epsilon_u\, \Delta_{u_3\text{L}}\, e^{\im\phi_u}\\
&& \Delta_{u_3\text{L}} \\
\end{pmatrix}
, \\
\Delta_{u \text{R}}^{5 \, \dagger} &=
\begin{pmatrix}
\Delta_{u_{12}\text{R}}^5 \\
& \Delta_{u_{12}\text{R}}^5 \\
&& \Delta_{u_3\text{R}}^5 \\
\end{pmatrix}
,\hspace{2em}
\Delta_{u \text{R}}^{6 \, \dagger} =
\begin{pmatrix}
\Delta_{u_{12}\text{R}}^6 \, e^{\im \phi_{u_{12}\text{R}}^6} \\
& \Delta_{u_{12}\text{R}}^6 \, e^{\im \phi_{u_{12}\text{R}}^6} \\
&& \Delta_{u_3\text{R}}^6 \, e^{\im \phi_{u_3\text{R}}^6} \\
\end{pmatrix}
, \\
\Delta_{d \text{L}} &=
\begin{pmatrix}
c_d\, \Delta_{d_1\text{L}} & -s_d\, \Delta_{d_2\text{L}}\, e^{\im\alpha_d} \\
s_d\, \Delta_{d_1\text{L}}\, e^{-\im\alpha_d} & c_d\, \Delta_{d_2\text{L}} &
\epsilon_d\, \Delta_{d_3\text{L}}\, e^{\im\phi_d}\\
&& \Delta_{d_3\text{L}} \\
\end{pmatrix}
, \\
\Delta_{d \text{R}}^{5 \, \dagger} &= \begin{pmatrix}
\Delta_{d_{12}\text{R}}^5 \\
& \Delta_{d_{12}\text{R}}^5 \\
&& \Delta_{d_3\text{R}}^5 \\
\end{pmatrix}
,\hspace{2em}
\Delta_{d \text{R}}^{6 \, \dagger} =
\begin{pmatrix}
\Delta_{d_{12}\text{R}}^6 \, e^{\im \phi_{d_{12}\text{R}}^6} \\
& \Delta_{d_{12}\text{R}}^6 \, e^{\im \phi_{d_{12}\text{R}}^6} \\
&& \Delta_{d_3\text{R}}^6 \, e^{\im \phi_{d_3\text{R}}^6} \\
\end{pmatrix},
\end{align} \label{eq:FlavourSpurions}
\end{subequations}
where all unphysical parameters have been rotated away, such that all parameters are real numbers.
Note that the relative phases between $\Delta^5_\text{R}$ and $\Delta^6_\text{R}$ are not fixed and appear as free parameters.

\section{Experimental analyses directly searching for heavy resonances}

\subsection{Searches for fermionic resonances}\label{app:fermion}
\begin{table}[H]
\centering
\resizebox{!}{238pt}{
\renewcommand{\arraystretch}{1.15}
\begin{tabular}{llllll}
\hline
Decay &Experiment &$\sqrt{s}$ [TeV] &Lum. [fb$^{-1}$] &Analysis &\\
\hline
$Q \to jZ$
&CDF	&1.96	&1.055 	&		&\cite{Aaltonen:2007je}\\%
\hline
\multirow{2}{*}{$Q \to jW$}
&ATLAS	&7	&1.04	&EXOT-2011-28	&\cite{Aad:2012bt}\\%
&CDF	&1.96	&4.6 	&		&\cite{CDF-PUB-TOP-PUBLIC-10110}\\%
\hline
\multirow{2}{*}{$Q \to qW$}
&CMS	&8	&19.7	&B2G-12-017	&\cite{CMS:2014dka}\\%
&ATLAS	&8	&20.3	&EXOT-2014-10	&\cite{Aad:2015tba}\\%
\hline
\multirow{5}{*}{$Q \to bW$}
&CMS	&7	&5	&EXO-11-050	&\cite{CMS:2012ab}\\%
&CMS	&7	&5	&EXO-11-099	&\cite{Chatrchyan:2012vu}\\%
&ATLAS	&7	&4.7	&EXOT-2012-07	&\cite{ATLAS:2012qe}\\%
&ATLAS	&8	&20.3	&CONF-2015-012	&\cite{ATLAS:2015dka}\\%
&CMS	&8	&19.7	&B2G-12-017	&\cite{CMS:2014dka}\\%
\hline
$Q \to tW$
&CMS	&7 	&5	&B2G-12-004	&\cite{Chatrchyan:2012af}\\
\hline
\multirow{3}{*}{$U \to tH$}
&CMS	&8	&19.7	&B2G-13-005	&\cite{Khachatryan:2015oba}\\%
&ATLAS	&13	&3.2	&CONF-2016-013	&\cite{ATLAS-CONF-2016-013}\\
&CMS	&13	&2.6	&PAS-B2G-16-011	&\cite{CMS:2016dmr}\\%
\hline
\multirow{4}{*}{$U \to tZ$}
&CMS	&7	&5	&B2G-12-004	&\cite{Chatrchyan:2012af}\\%
&CMS	&7	&1.1	&EXO-11-005	&\cite{Chatrchyan:2011ay}\\%
&CMS	&8	&19.7	&B2G-13-005	&\cite{Khachatryan:2015oba}\\%
&ATLAS	&13	&14.7	&CONF-2016-101	&\cite{ATLAS:2016qlg}\\%
\hline
\multirow{2}{*}{$U \to bW$}
&CMS	&8	&19.7	&B2G-13-005	&\cite{Khachatryan:2015oba}\\%
&ATLAS	&13	&14.7	&CONF-2016-102	&\cite{ATLAS:2016cuv}\\%
\hline
\multirow{4}{*}{$D \to bH$}
&ATLAS	&8	&20.3	&CONF-2015-012	&\cite{ATLAS:2015dka}\\%
&CMS	&8	&19.8	&B2G-12-019	&\cite{CMS:2012hfa}\\%
&CMS	&8	&19.5	&B2G-13-003	&\cite{CMS:2013una}\\%
&CMS	&8	&19.7	&B2G-14-001	&\cite{CMS:2014afa}\\%
\hline
\multirow{3}{*}{$D \to bZ$}
&CMS	&7	&5	&EXO-11-066	&\cite{CMS:2012jwa}\\%
&CMS	&8	&19.5	&B2G-13-003	&\cite{CMS:2013una}\\%
&CMS	&8	&19.7	&B2G-13-006	&\cite{Khachatryan:2015gza}\\%
\hline
\multirow{4}{*}{$D \to tW$}
&ATLAS	&8	&20.3	&EXOT-2013-16	&\cite{Aad:2015gdg}\\%
&CMS	&8	&19.5	&B2G-13-003	&\cite{CMS:2013una}\\%
&CMS	&8	&19.7	&B2G-13-006	&\cite{Khachatryan:2015gza}\\%
&CDF	&1.96	&2.7	&		&\cite{Aaltonen:2009nr}\\%
\hline
\multirow{3}{*}{$Q_{5/3} \to tW$}
&ATLAS	&8	&20.3	&EXOT-2014-17	&\cite{Aad:2015mba}\\%
&CMS	&8	&19.6	&B2G-12-012	&\cite{CMS:vwa}\\%
&CMS	&13	&2.2	&PAS-B2G-15-006	&\cite{CMS:2015alb}\\%
\hline
\end{tabular}
}
\caption{Experimental analyses included in our numerics for heavy quark
partner decay. $Q$ stands for any quark partner where the decay in question is
allowed by electric charges, $j$ stands for a light quark or $b$ jet, and $q$
for a light quark jet.}
\label{tab:exp_quark_res}
\end{table}

\subsection{Searches for bosonic resonances}\label{app:bosons}

\begin{table}[H]
\centering
\resizebox{!}{253pt}{
\renewcommand{\arraystretch}{1.15}
\begin{tabular}{llllll}
\hline
Decay &Experiment &$\sqrt{s}$ [TeV] &Lum. [fb$^{-1}$] &Analysis &\\
\hline
\multirow{6}{*}{$\eta \to hh$}
&CMS	&8	&19.7	&PAS-EXO-15-008		&\cite{CMS:2015zug}\\%
&ATLAS	&13	&3.2	&EXOT-2015-11		&\cite{Aaboud:2016xco}\\%
&CMS	&13	&2.3	&PAS-HIG-16-002		&\cite{CMS:2016tlj}\\%
&CMS	&13	&2.7	&PAS-B2G-16-008		&\cite{CMS:2016pwo}\\%
&CMS	&13	&12.9	&PAS-HIG-16-029		&\cite{CMS:2016knm}\\%
&CMS	&13	&2.7	&PAS-HIG-16-032		&\cite{CMS:2016vpz}\\%
\hline
\multirow{5}{*}{$\eta \to ZZ$}
&ATLAS	&13	&13.3	&CONF-2016-056		&\cite{ATLAS:2016bza}\\%
&ATLAS	&13	&14.8	&CONF-2016-079		&\cite{ATLAS:2016oum}\\%
&ATLAS	&13	&13.2	&CONF-2016-082		&\cite{ATLAS:2016npe}\\%
&CMS	&13	&12.9	&PAS-HIG-16-033		&\cite{CMS:2016ilx}\\%
&CMS	&13	&2.7	&PAS-B2G-16-010		&\cite{CMS:2016tio}\\%
\hline
\multirow{6}{*}{$\eta \to W^+W^-$}
&ATLAS	&8	&20.3	&EXOT-2013-01$^*$	&\cite{Aad:2015ufa}\\%
&CMS	&8	&19.7	&EXO-13-009$^*$		&\cite{Khachatryan:2014gha}\\%
&ATLAS	&13	&13.2	&CONF-2016-062$^*$	&\cite{ATLAS:2016cwq}\\%
&ATLAS	&13	&13.2	&CONF-2016-074		&\cite{ATLAS:2016kjy}\\%
&CMS	&13	&2.3	&PAS-HIG-16-023		&\cite{CMS:2016jpd}\\%
\hline
\multirow{2}{*}{$\eta \to \gamma\gamma$}
&ATLAS	&13	&15.4	&CONF-2016-059		&\cite{ATLAS:2016eeo}\\%
&CMS	&13	&16.2	&PAS-EXO-16-027		&\cite{CMS:2016crm}\\%
\hline
\multirow{5}{*}{$\eta \to Z\gamma$}
&ATLAS	&13	&13.3	&CONF-2016-044		&\cite{ATLAS:2016lri}\\%
&ATLAS	&13	&3.2	&EXOT-2016-02		&\cite{Aaboud:2016trl}\\%
&CMS	&13	&19.7	&PAS-EXO-16-025		&\cite{CMS:2016mvc}\\%
&CMS	&13	&12.9	&PAS-EXO-16-034		&\cite{CMS:2016pax}\\%
&CMS	&13	&12.9	&PAS-EXO-16-035		&\cite{CMS:2016cbb}\\%
\hline
\multirow{4}{*}{$\eta \to e^+e^-/\mu^+\mu^-$}
&ATLAS	&8	&20.3	&EXOT-2012-23$^*$	&\cite{Aad:2014cka}\\%
&CMS	&8	&20.6	&EXO-12-061$^*$		&\cite{Khachatryan:2014fba}\\%
&ATLAS	&13	&13.3	&CONF-2016-045$^*$	&\cite{ATLAS:2016cyf}\\%
&CMS	&13	&12.4	&PAS-EXO-16-031$^*$	&\cite{CMS:2016abv}\\%
\hline
\multirow{3}{*}{$\eta \to \tau^+\tau^-$}
&ATLAS	&8	&19.5	&EXOT-2014-05$^*$	&\cite{Aad:2015osa}\\%
&CMS	&8	&19.7	&EXO-12-046$^*$		&\cite{CMS:2015ufa}\\%
&CMS	&13	&2.2	&PAS-EXO-16-008$^*$	&\cite{CMS:2016zxk}\\%
&ATLAS	&13	&13.3	&CONF-2016-085		&\cite{ATLAS:2016fpj}\\%
&CMS	&13	&2.3	&PAS-HIG-16-006		&\cite{CMS:2016pkt}\\%
\hline
\multirow{4}{*}{$\eta \to t\bar t$}
&ATLAS	&8	&20.3	&CONF-2015-009$^*$	&\cite{ATLAS:2015aka}\\%
&CMS	&8	&19.7	&B2G-13-008$^*$		&\cite{Khachatryan:2015sma}\\%
&CMS	&13	&2.6	&PAS-B2G-15-002$^*$	&\cite{CMS:2016zte}\\%
&CMS	&13	&2.6	&PAS-B2G-15-003$^*$	&\cite{CMS:2016ehh}\\%
\hline
$\eta \to b\bar{b}$
&CMS	&13	&2.69	&PAS-HIG-16-025		&\cite{CMS:2016ncz}\\%
\hline
$\eta \to qq$
&CMS	&13	&12.9	&PAS-EXO-16-032$^*$	&\cite{CMS:2016wpz}\\%
\hline
$\eta \to gg$
&CMS	&13	&12.9	&PAS-EXO-16-032		&\cite{CMS:2016wpz}\\%
\hline
\multirow{2}{*}{$\eta \to jj$}
&ATLAS	&13	&3.6	&EXOT-2015-02$^*$	&\cite{ATLAS:2015nsi}\\%
&CMS	&13	&2.4	&EXO-15-001$^*$		&\cite{Khachatryan:2015dcf}\\%
\hline
\end{tabular}
}
\caption{Experimental analyses included in our numerics for $\eta$ decay.
The analyses marked with~$^*$ are actually searches for neutral vector resonances. Since for many channels there are no dedicated analyses searching for a neutral scalar resonance and the bounds should be similar, we include the spin-1 analyses in our numerics for $\eta$ decay.}
\label{tab:exp_scalar_res}
\end{table}

\begin{table}[H]
\centering
\resizebox{!}{287pt}{
\renewcommand{\arraystretch}{1.15}
\begin{tabular}{llllll}
\hline
Decay &Experiment &$\sqrt{s}$ [TeV] &Lum. [fb$^{-1}$] &Analysis &\\
\hline
\multirow{5}{*}{$\rho^{\pm} \to W^{\pm}h$}
&ATLAS	&8	&20.3	&EXOT-2013-23	&\cite{Aad:2015yza}\\%
&CMS	&8	&19.7	&EXO-14-010	&\cite{CMS:2015gla}\\%
&ATLAS	&13	&3.2	&EXOT-2015-18	&\cite{Aaboud:2016lwx}\\%
&ATLAS	&13	&13.3	&CONF-2016-083	&\cite{ATLAS:2016kxc}\\%
&CMS	&13	&2.17	&PAS-B2G-16-003	&\cite{CMS:2016dzw}\\%
\hline
\multirow{9}{*}{$\rho^{\pm} \to W^{\pm}Z$}
&ATLAS	&8	&20.3	&EXOT-2013-01	&\cite{Aad:2015ufa}\\%
&ATLAS	&8	&20.3	&EXOT-2013-07	&\cite{Aad:2014pha}\\%
&ATLAS	&8	&20.3	&EXOT-2013-08	&\cite{Aad:2015owa}\\%
&CMS	&8	&19.7	&EXO-12-024	&\cite{Khachatryan:2014hpa}\\%
&ATLAS	&13	&15.5	&CONF-2016-055	&\cite{ATLAS:2016yqq}\\%
&ATLAS	&13	&13.2	&CONF-2016-062	&\cite{ATLAS:2016cwq}\\%
&ATLAS	&13	&13.2	&CONF-2016-082	&\cite{ATLAS:2016npe}\\%
&CMS	&13	&2.2	&PAS-EXO-15-002	&\cite{CMS:2015nmz}\\%
&CMS	&13	&12.9	&PAS-B2G-16-020	&\cite{CMS:2016pfl}\\%
\hline
\multirow{4}{*}{$\rho^{\pm} \to tb$}
&CMS	&8	&19.5	&B2G-12-010	&\cite{Chatrchyan:2014koa}\\%
&CMS	&8	&19.7	&B2G-12-009	&\cite{Khachatryan:2015edz}\\%
&CMS	&13	&2.55	&PAS-B2G-16-009	&\cite{CMS:2016ude}\\%
&CMS	&13	&12.9	&PAS-B2G-16-017	&\cite{CMS:2016wqa}\\%
\hline
\multirow{2}{*}{$\rho^{\pm} \to \tau^{\pm}\nu$}
&CMS	&8	&19.7	&EXO-12-011	&\cite{Khachatryan:2015pua}\\%
&CMS	&13	&2.3	&PAS-EXO-16-006	&\cite{CMS:2016ppa}\\%
\hline
\multirow{3}{*}{$\rho^{\pm} \to e^{\pm}\nu/\mu^{\pm}\nu$}
&ATLAS	&7	&4.7	&EXOT-2012-02	&\cite{Aad:2012dm}\\%
&ATLAS	&13	&13.3	&CONF-2016-061	&\cite{ATLAS:2016ecs}\\%
&CMS	&13	&2.2	&PAS-EXO-15-006	&\cite{CMS:2015kjy}\\%
\hline
$\rho^{\pm} \to jj$
&ATLAS	&13	&3.6	&EXOT-2015-02	&\cite{ATLAS:2015nsi}\\%
\hline
\multirow{3}{*}{$\rho^0 \to W^+W^-$}
&ATLAS	&8	&20.3	&EXOT-2013-01	&\cite{Aad:2015ufa}\\%
&CMS	&8	&19.7	&EXO-13-009	&\cite{Khachatryan:2014gha}\\%
&ATLAS	&13	&13.2	&CONF-2016-062	&\cite{ATLAS:2016cwq}\\%
\hline
\multirow{6}{*}{$\rho^0 \to Zh$}
&ATLAS	&8	&20.3	&EXOT-2013-23	&\cite{Aad:2015yza}\\
&CMS	&8	&19.7	&EXO-13-007	&\cite{Khachatryan:2015ywa}\\
&ATLAS	&13	&3.2	&EXOT-2015-18	&\cite{Aaboud:2016lwx}\\
&ATLAS	&13	&3.2	&CONF-2015-074	&\cite{ATLAS-CONF-2015-074}\\
&ATLAS	&13	&13.3	&CONF-2016-083	&\cite{ATLAS:2016kxc}\\
&CMS	&13	&2.17	&PAS-B2G-16-003	&\cite{CMS:2016dzw}\\
\hline
$\rho^0 \to W^+W^-/Zh$
&CMS	&13	&2.2	&PAS-B2G-16-007	&\cite{CMS:2016wev}\\%
\hline
\multirow{4}{*}{$\rho^0 \to e^+e^-/\mu^+\mu^-$}
&ATLAS	&8	&20.3	&EXOT-2012-23	&\cite{Aad:2014cka}\\%
&CMS	&8	&20.6	&EXO-12-061	&\cite{Khachatryan:2014fba}\\%
&ATLAS	&13	&13.3	&CONF-2016-045	&\cite{ATLAS:2016cyf}\\%
&CMS	&13	&12.4	&PAS-EXO-16-031	&\cite{CMS:2016abv}\\%
\hline
\multirow{3}{*}{$\rho^0 \to \tau^+\tau^-$}
&ATLAS	&8	&19.5	&EXOT-2014-05	&\cite{Aad:2015osa}\\%
&CMS	&8	&19.7	&EXO-12-046	&\cite{CMS:2015ufa}\\%
&CMS	&13	&2.2	&PAS-EXO-16-008	&\cite{CMS:2016zxk}\\%
\hline
\multirow{4}{*}{$\rho^0/\rho_G \to t\bar t$}
&ATLAS	&8	&20.3	&CONF-2015-009	&\cite{ATLAS:2015aka}\\%
&CMS	&8	&19.7	&B2G-13-008	&\cite{Khachatryan:2015sma}\\%
&CMS	&13	&2.6	&PAS-B2G-15-002	&\cite{CMS:2016zte}\\%
&CMS	&13	&2.6	&PAS-B2G-15-003	&\cite{CMS:2016ehh}\\%
\hline
\multirow{2}{*}{$\rho^0/\rho_G \to jj$}
&ATLAS	&13	&3.6	&EXOT-2015-02	&\cite{ATLAS:2015nsi}\\%
&CMS	&13	&2.4	&EXO-15-001	&\cite{Khachatryan:2015dcf}\\%
\hline
$\rho^0/\rho_G \to qq$
&CMS	&13	&12.9	&PAS-EXO-16-032	&\cite{CMS:2016wpz}\\%
\hline
\end{tabular}
}
\caption{Experimental analyses included in our numerics for heavy vector
resonance decay.}
\label{tab:exp_vector_res}
\end{table}

\newpage
\bibliographystyle{JHEP}
\bibliography{bibliography}

\end{document}